\documentclass[12pt]{iopart}

\usepackage{epsfig}
\usepackage{amssymb}
\usepackage{iopams}
\usepackage{comment}
\usepackage{braket}
\usepackage{enumerate}
\newcommand{\be}{\begin{equation}}
\newcommand{\ee}{\end{equation}}
\newcommand{\ba}{\begin{aligned}}
\newcommand{\ea}{\end{aligned}}
\renewcommand{\tr}[2]{\mathrm{tr}_{#1}\bigl[#2\bigr]}

\newcommand{\bea}{\begin{eqnarray}}
\newcommand{\eea}{\end{eqnarray}}
\def\nn{\nonumber\\}

\eqnobysec

\begin{document}
\title{On Conservation Laws, Relaxation and Pre-relaxation after a Quantum Quench}  
\author{Maurizio Fagotti}
\address{The Rudolf Peierls Centre for Theoretical Physics,
    Oxford University, Oxford, OX1 3NP, United Kingdom}
\ead{Maurizio.Fagotti@physics.oc.ac.uk}
\begin{abstract}
We consider the time evolution following a quantum quench in spin-1/2 chains. It is well known that local conservation laws constrain the dynamics and, eventually, the stationary behavior of local observables.
We show that some widely studied models, like the quantum XY model, possess extra families of local conservation laws in addition to the translation invariant ones. As a consequence, the additional charges must be included in the generalized Gibbs ensemble that describes the stationary properties. The effects go well beyond a simple redefinition of the stationary state. The time evolution of a non-translation invariant state under a (translation invariant) Hamiltonian with a perturbation that weakly breaks the hidden symmetries underlying the extra conservation laws exhibits pre-relaxation. In addition, in the limit of small perturbation, the time evolution following pre-relaxation can be described by means of a time-dependent generalized Gibbs ensemble.

\end{abstract}
\maketitle
\section{Introduction}\label{s:intro}

The non-equilibrium time evolution under a Hamiltonian with local interactions is the simplest example of out-of-equilibrium dynamics. 
Pioneering experiments~\cite{kww-06,tetal-11,cetal-12,getal-11,shr-12}  stimulated a renewed theoretical interest in the subject, which is now one of the most promising areas of research,  spreading from the most fundamental aspects of quantum physics to the most advanced techniques for experiment arrangements.  

If the system is initially prepared in the ground state of a globally different Hamiltonian, the protocol is usually called global (quantum, sudden) quench. One of the most interesting aspects of quench dynamics is the time relaxation of local degrees of freedom. At first, this might be perceived as a counterintuitive effect, being the time evolution unitary and relaxation generally associated with dissipative processes. Local relaxation is however possible in infinite systems, since information can flow towards infinity without ever coming back.

In the last few years several descriptions of the stationary properties have been proposed. Essentially,  they are based on replacing the density matrix (the projector on the time dependent state) with a stationary state, which can be either mixed, as the (block) diagonal ensemble~\cite{BKL_PRL10,CCR_PRL11, F:13a} and the (generalized) Gibbs ensemble~\cite{FE_13b,FCEC-13, BS_PRL08,CEF,FE_13a,CC_JStatMech07,IC_PRA09,FM_NJPhys10,EEF:12,Pozsgay_JStatMech11,CIC_PRE11,CK_PRL12,MC_NJPhys12,Pozsgay:13a,CSC:13b,KCC:13a,CE_10, mc-12b,m-13, KSCCI}, or pure, as a representative state~\cite{CE_PRL13,nwbc-13}. In the thermodynamic limit, all these states share the same reduced density matrices. 

In this respect a fundamental question is how much and what kind of information  we need in order to construct a stationary state with the correct reduced density matrices.

It is widely believed that in a generic system the stationary properties can be described by a Gibbs ensemble with effective temperature fixed by the energy conservation~\cite{nonint} (or any other constraint that involves only local degrees of freedom). Despite the practical difficulties in dealing with that ensemble, this is a very simple representation written in terms of the Hamiltonian and a real parameter\footnote{In contrast, the diagonal ensemble is extremely complicated, as there may be accidental degeneracies which are out of our control.}.

Integrable models make an exception: the representation of the stationary state involves an infinite number of parameters that must be somehow 
fixed.  
In spin chains this can be easily understood by considering that an integrable model possesses an infinite number of conservation laws in involution~\cite{Korepinbook,GM-94} $[H_n,H_m]=0$ ($H_1\equiv H$ is the Hamiltonian)  of the form
\be\label{eq:Hn}
H_n=\sum_\ell h_\ell^{(n)}\, ,
\ee   
where, for fixed $n$, $h_\ell^{(n)}$ is an operator that acts nontrivially only on a finite subsystem that includes the site $\ell$ and has length independent of $\ell$. These are generally called \emph{local} conservation laws, although `local' clearly refers to the operators $h_\ell^{(n)}$. They constrain the  stationary state $\bar\rho$ resulting from the time evolution of $\ket{\Psi_0}$ because
\be\label{eq:iom}
\fl\qquad\qquad\braket{\Psi_0|H_n|\Psi_0}=
\lim_{t\rightarrow\infty}\sum_\ell \braket{\Psi_0| e^{i H t} h_\ell^{(n)} e^{-i H t}|\Psi_0}=\sum_\ell \tr{}{\bar \rho\  h_\ell^{(n)} }\, ,
\ee 
where in the first identity we used that $H_n$ is conserved and in the second we assumed that at late times finite subsystems can be described by $\bar\rho$.

We notice that \eref{eq:iom} holds true for a larger class of charges in which the operators $h_\ell^{(n)}$ are not local but have tails that decay sufficiently fast  for the second identity to be satisfied. We qualify them as `quasi-local' and for the sake of simplicity we shall restrict to exponentially localized operators~\cite{IP-13}. 

In the following we will call \emph{`maximal'} a set of local charges that commutes with a (quasi-)local conservation law only if the latter is a linear combination of its elements.

One of the most ``physical'' representations of the stationary state is the generalized Gibbs ensemble~\cite{GGE}, which is the mixed state with maximal entanglement entropy under the constraints of a maximal set of local conservation laws \eref{eq:Hn}:
\be\label{eq:GGE}
\rho_{\rm GGE}=\frac{1}{Z}e^{-\sum_{n} \lambda_n H_n}\, .
\ee
As shown in \cite{FE_13a} and partially justified by \eref{eq:iom}, dropping a single local conservation law in \eref{eq:GGE} produces local effects. On the other hand, disregarding a nonlocal conservation law (still linear combination of $H_n$, but clearly involving an infinite number of local charges) generally gives rise to an equivalent representation (in noninteracting models this can be easily shown for the removal of a mode occupation number).

This description is therefore based on two assumptions: 
\begin{enumerate}[(a)]
\item \label{hyp:a} (GGE hypothesis) The stationary state can be represented by a statistical ensemble that has maximal entanglement entropy under the constraints of the local conservation laws;
\item \label{hyp:b} Any (quasi-)local conservation law can be written as a linear combination of the maximal set \eref{eq:Hn} of local charges  in involution. 
\end{enumerate}
While the validity of assumption~\eref{hyp:a} is still under investigation and has been attracting the attention of a vast community of physicists, it seems that hypothesis~\eref{hyp:b} is generally assumed to be true quite implicitly, perhaps without a real perception of the problem. 
However, if there is an independent local conservation law $Q$, the corresponding integral of motion can be written as in \eref{eq:iom}, giving a further constraint to the stationary state. 
This is a topical issue: generally it is extremely difficult to address questions like \eref{hyp:b} and recently new independent conservation laws have been found for the XXZ model~\cite{prosen}.

In order to gain some insight into that problem, we focus on the simplest cases in which maximal sets of charges in involution can be clearly identified.  

We will then call \emph{`superintegrable'} a model that has extra families of \emph{local} conservation laws in addition to a maximal set in involution (which accounts for integrability)\footnote{
This definition of `superintegrability' fully relies on locality and is therefore more restrictive than others~\cite{super};  in fact, it is aimed at emphasizing the impact on local observables in quantum many-body systems out of equilibrium.}. 
Because of the additional constraints, the generalized Gibbs ensemble may be different from the one shown in \eref{eq:GGE}.
However, this does not invalidate assumption~\eref{hyp:a}.

The first aim of this paper is to show that some of the paradigms of  non-equilibrium dynamics exhibit additional local conservation laws. One of them is the celebrated quantum XY model in the absence of a magnetic field, which is widely studied~\cite{FC:2008,bpgda-10}, especially in the isotropic limit, known as  XX model. 
We construct an extra family of local charges and discuss the effects on the stationary state after the quench. 

Besides describing such special cases, we show that the additional integrals of motion affect also the non-equilibrium time evolution of integrable models close to superintegrability points. 
Subsystems can experience a pre-relaxation time window which can be approximately described by a generalized Gibbs ensemble.  Similar behavior has been recently observed after quantum quenches in nonintegrable models and in open quantum systems~\cite{getal-11,EKMR-13,pre-therm1,pre-therm2,pre-therm3,pre-therm4,pre-therm5,pre-therm6,pre-therm7,pre-therm8,MS-12}, in which cases it was named `pre-thermalization'.

Interestingly, we find that in the limit of weak perturbation a \emph{time-dependent GGE} can be used to describe the time evolution following pre-relaxation.
To the best of our knowledge, this is the first comprehensive description proposed for the relaxation process of a system that exhibits pre-relaxation. 

We discuss two relevant examples: a global quench in the quantum XY model with a small magnetic field and the non-equilibrium evolution of the Majumdar-Ghosh ground state under the Hamiltonian of the XXZ spin-1/2 chain in the Ising limit of large anisotropy. 
In the latter case the Hamiltonian is interacting, however the particular choice of the initial state allows us to obtain some analytic results. 

\subsection*{Organization of the manuscript}
The paper is organized as follows: \Sref{s:deg} provides an overview of the relation between accidental degeneracy and appearance of additional conservation laws and a discussion of the effects on the stationary state after the quench. \Sref{s:lcl} is dedicated to the construction of the local conservation laws in noninteracting spin-1/2 chains; 
we present a self-consistent and rather detailed analysis, which the uninterested reader might skip.   
In Sections \ref{s:time} and \ref{s:time1} it is considered the time evolution after quantum quenches 
of integrable models close to superintegrability points. \Sref{s:conc} contains our conclusions.

\section{Accidental degeneracies and Extra Families of Local Conservation Laws}\label{s:deg}
We start off with a question: Let $H$ be the noninteracting Hamiltonian
\be\label{eq:Hfree}
H=\sum_{k}\varepsilon(k)\Bigl(b^\dag_k b_k^{\phantom{\dag}}-\frac{1}{2}\Bigr)\, ,
\ee
where $b^\dag_k$ are noninteracting spinless fermions ($\{b^\dag_k,b_{p}\}=\delta_{k p}$, $\{b_k,b_{p}\}=0$) and $\varepsilon(k)$ is the dispersion relation; what is the most general quadratic operator commuting with~$H$? 

One might be tempted to answer: a linear combination of the mode occupation numbers $b^\dag_k b_k^{\phantom{\dag}}$. In fact, the answer depends on the dispersion relation. For example, if the latter is flat, any operator of the form $b^{\dag}_k b_p^{\phantom{\dag}}$ commutes with $H$ (indeed \mbox{$[b^{\dag}_k b_p^{\phantom{\dag}},b^{\dag}_k b_k^{\phantom{\dag}}+b^{\dag}_p b_p^{\phantom{\dag}}]=0$}). We could therefore define new fermions of the form \mbox{$\alpha_\sigma(k) b^\dag_k+\beta_\sigma(k) b^\dag_{\sigma(k)}$}, where $\sigma$ is some permutation operator, whose occupation numbers are independent of $b^\dag_k b^{\phantom{\dag}}_k$ and commute with the Hamiltonian.
It is noteworthy that in the finite system the new occupation numbers provide a complete set of quantum numbers. 

More in general, degeneracies of the one-particle spectrum ($\varepsilon_k=\varepsilon_p$ for two distinct momenta $k$ and $p$) allow us to define independent sets of (quadratic) conservation laws that commute between each other. 
Let us notice that such accidental degeneracy is the rule, indeed the dispersion relation of a noninteracting translation invariant spin-1/2 Hamiltonian with local interactions is generally the absolute value of a smooth periodic function, so the energy $\varepsilon_k$ of every mode $k$ with a nonzero velocity $\varepsilon_k^\prime$ is at least double degenerate.

The arbitrariness in the definition of the fermions that diagonalize the Hamiltonian is just a particular case of the fact that eigenvectors of degenerate eigenspaces are not univocally defined. It becomes however a powerful observation when $H$ is the post-quench Hamiltonian and we wish to describe the late time stationary behavior. 

For the sake of concreteness we consider the Hamiltonian of the XY model without magnetic field
\be\label{eq:HXY0}
H=J\sum_\ell \Bigl[\frac{1+\gamma}{4} \sigma_\ell^x\sigma_{\ell+1}^x+\frac{1-\gamma}{4} \sigma_\ell^y\sigma_{\ell+1}^y\Bigr]\, ,
\ee
which includes the (critical) XX model as a special case $\gamma=0$. The spin operators $\sigma_\ell^\alpha$ act on the site $\ell$ as Pauli matrices. The Hamiltonian is quadratic in the Jordan-Wigner fermions
\be
c^\dag_\ell=\left(\prod_{j<\ell}\sigma_j^z\right) \frac{\sigma_\ell^x+i\sigma_\ell^y}{2}
\ee 
and can be easily mapped to \eref{eq:Hfree}.  
A set of local conservation laws can be obtained \emph{e.g.} with the method discussed  in \cite{FE_13a}, which results in a maximal set of (translation invariant) charges of the form (some details can be found in the next section)
\be\label{eq:claws}
I^+_n=\sum_{k}\cos((n-1) k)\varepsilon(k)b^\dag_k b_k^{\phantom{\dag}}\, ,\qquad I^-_n=\sum_{k}\sin(n k)b^\dag_k b_k^{\phantom{\dag}}\, .
\ee
These involve only spin operators acting (nontrivially) on $n+1$ neightboring sites. Let us notice that, for any finite periodic chain, $I^\pm_n$ produce a complete set of quantum numbers, provided that $n$ is allowed to assume values comparable with the chain length.

Having a maximal set of local conservation laws in involution (or, equivalently, mode occupation numbers), we might expect to be able to construct the generalized Gibbs ensemble corresponding to a generic initial state. In fact, if in the initial state one-site shift invariance is broken we are destined to failure. And the reason is that there are local conservation laws that break one-site shift invariance, that is to say there are integrals of motion that cannot be satisfied by an ensemble of the form \eref{eq:GGE} constructed with the charges~\eref{eq:claws}.  
\Eref{eq:HXY0} is indeed the Hamiltonian of a \emph{superintegrable} model, as defined in \Sref{s:intro}. For example, the operator
\be\label{eq:new}
\sum_\ell \Bigl[\frac{1+\gamma}{4} \sigma_{2\ell}^x\sigma_{2\ell+1}^x+\frac{1-\gamma}{4} \sigma_{2\ell-1}^y\sigma_{2\ell}^y\Bigr]
\ee
commutes with the Hamiltonian~\eref{eq:HXY0} but is not a function of the charges~\eref{eq:claws} (this is straightforward, indeed \eref{eq:claws} are translation invariant while \eref{eq:new} is not). 

As it will become clear in the next section, we could still represent the generalized Gibbs ensemble in terms of (quasi-)local conservation laws in involution; however the set of charges would not be independent of the initial state!
This is rather unusual, as one generally expects that the relevant information about the initial state is encoded in the Lagrange multipliers $\lambda_n$ \eref{eq:GGE}.
In addition, analogous issues arise in the alternative approach in which the stationary properties are described by a representative state~\cite{CE_PRL13}: the set of quantum numbers required to define the state would depend in a nontrivial way on the particular set of charges (and, in turn, on the initial state). 

We prefer an alternative point of view, which has the advantage to fall into the standard definition of GGE, in the generic situation, and not to depend explicitly on the properties of the initial state, in the exceptional cases we are considering. 

We define the GGE  as the statistical ensemble $\rho_{\rm GGE^*}$ with maximal entanglement entropy $-\tr{}{\rho_{\rm GGE^*}\log\rho_{\rm GGE^*}}$ under the constraints of the \emph{linearly independent local} conservation laws $Q_j$. By assuming that the maximum is a stationary point we end up with the condition
\be
-\tr{}{\delta\rho\log\rho_{\rm GGE^*}}-\tr{}{\delta\rho(\lambda_0+\sum_{j=1}\lambda_jQ_j)}=0
\ee
which is satisfied for a generic variation $\delta\rho$ if
\be\label{eq:GGEimp}
\rho_{\rm GGE^*}=\frac{e^{-\sum_{j=1}\lambda_jQ_j}}{Z}\, .
\ee 
So far, we never used that the charges $Q_j$ commute with one another; they must only commute with the Hamiltonian $H\equiv Q_1$. 
However, the commutator of two operators with local densities has local density as well, therefore the Jacobi identity implies that the operators $Q_j$ satisfy a closed algebra:
\be\label{eq:n-a}
\fl\qquad{}[H,i [Q_j,Q_k]]=-i [Q_j,[Q_k,H]]-i [Q_k,[H,Q_j]]=0\Longrightarrow i [Q_j,Q_k]=f_{j k \ell} Q_\ell\, .
\ee
In a superintegrable model the structure constants $f_{j k \ell}$ are not trivially equal to zero; this is the reason why `superintegrability' is also addressed as `non-Abelian integrability'.  
We notice that, because of \eref{eq:n-a}, a single additional local conservation law is generally sufficient to produce an extra (potentially infinite) family of local charges. 

Finally, we notice that this approach allows us to use the same regularization introduced in \cite{FE_13a}, in which the GGE is defined as the limit of truncated generalized Gibbs ensembles constructed retaining a finite number of the most local conservation laws. 
The only difference is that now the set includes also the additional charges. After having ordered the conservation laws by their range, the truncated GGE is defined in an analogous way.

\section{Local charges of noninteracting translation invariant models}\label{s:lcl}
In this section we investigate the local conservation laws of noninteracting spin-1/2 chains. 
Let us consider a generic one-site shift invariant Hamiltonian\footnote{These are generally referred as `translation invariant' Hamiltonians, however in the following it will be important the number of sites needed to realize an elementary translation, so we prefer to use the term `shift invariant'.} with finite range and constructed with the building blocks
\be
\sum_\ell \sigma_\ell^z,\qquad \sum_\ell \sigma_\ell^{\rho}\mathcal S^z_{\ell,\ell+r}\sigma_{\ell+r}^{\rho^\prime} \qquad \rho,\rho^\prime\in\{x,y\}
\ee
where $\mathcal S^z_{\ell,\ell+n}=\prod_{j=\ell}^{\ell+n}\sigma_j^z$. Ignoring boundary terms, we can use the Jordan-Wigner transformation
\be\label{eq:JW}
a_\ell^{x,y}=\mathcal S^z_{1,\ell-1}\sigma_\ell^{x,y}
\ee
 to express the Hamiltonian as a quadratic form in the Majorana fermions $a_\ell^{x,y}$
\be
H=\frac{1}{4}\sum_{\ell,n}^{L}\left(\begin{array}{cc}a_{\ell}^x&a^{y}_{\ell}
\end{array} \right)\mathcal H^{(1)}_{\ell n}\left(\begin{array}{c}a_{n}^x\\a^{y}_{n}
\end{array}\right)\, ,
\ee
where $L$ is the chain length (which eventually will be sent to infinity) and $\mathcal H_{l n}^{(1)}$ is a block-circulant matrix whose elements can be written in terms of a 2-by-2 matrix $\mathcal H^{(1)}(k)$, generally called \emph{symbol}, as follows
\be\label{eq:anti}
\mathcal H^{(1)}_{l n}=\frac{1}{L}\sum_{k} e^{-i(n-\ell)k}\mathcal H^{(1)}(k)\, ,\qquad e^{i k L}=1\, .
\ee
This is only one of the block-circulant representations of the Hamiltonian. More in general, we can work with $(2n)$-by-$(2n)$ block-circulant matrices, which differ in the number of fermions that are associated with a single block.
For example, for $n=2$ the Hamiltonian reads as
\be\label{eq:H4}
H=\frac{1}{4}\sum_{\ell,n}^{L/2}\left(\begin{array}{cccc}a_{2\ell-1}^x&a^{y}_{2\ell-1}&a_{2\ell}^x&a_{2\ell}^y
\end{array} \right)\mathcal H^{(2)}_{\ell n}\left(\begin{array}{c}a_{2n-1}^x\\a^{y}_{2n-1}\\a_{2n}^x\\a_{2n}^y
\end{array}\right)\, ,
\ee
and, in general, the symbol $\mathcal H^{(n)}(k)$ of the block-circulant matrix is defined by the equation 
\be
\mathcal H^{(n)}_{l m}=\frac{n}{L}\sum_{k} e^{-i(m-\ell)k}\mathcal H^{(n)}(k)\, ,\qquad e^{i k L/n}=1\, .
\ee

Since we are considering one-site shift invariant Hamiltonians, the symbol of the $n$-site representation is in a simple relation with $\mathcal H^{(1)}(k)$; in particular, for $n=2$
we find
\be\label{eq:from1to2}
\mathcal H^{(2)}(k)=\frac{1+\tau^x e^{i\frac{k}{2}\tau^z}}{2}\otimes\mathcal H^{(1)}(k/2)+\frac{1-\tau^x e^{i\frac{k}{2}\tau^z}}{2}\otimes\mathcal H^{(1)}(k/2+\pi)
\ee
where $\tau^\alpha$ are Pauli matrices.

We notice that there is an ambiguity in the definition of the symbol, which is removed by requiring
\be\label{eq:conditions}
{\mathcal H^{(n)}}^\dag(k)=\mathcal H^{(n)}(k)\qquad {\mathcal H^{(n)}}^t(k)=-\mathcal H^{(n)}(-k)\, ,
\ee
which reflect Hermiticity and the algebra of the Majorana fermions ($\{a_i^{x(y)},a_j^{x(y)}\}=2\delta_{i j}$, $\{a_i^x,a_j^y\}=0$).

Briefly, the $n$-site representation implicitly defines $n$ species of fermions with dispersion relation given by the positive eigenvalues of $\mathcal H^{(n)}(k)$. At fixed momentum~$k$, if the dispersion relations are distinct, the (Bogoliubov) fermions that annihilate the ground state and diagonalize the Hamiltonian are defined up to a phase. If instead there is some degeneracy, the species of fermions can be mixed, as described at the beginning of \Sref{s:deg}.

In conclusion, degeneracy in the symbol indicates that there are further quadratic operators commuting with the Hamiltonian, besides a complete set of occupation numbers.
However, the problem of whether these are local is far more complicated. 

\subsection{Symbols of local conservation laws}
The one-to-one correspondence between shift-invariant noninteracting fermionic operators ($H$) and the associated block-circulant matrices ($\mathcal H^{(n)}$) is so powerful that we can generally reduce operations in the fermionic Hilbert space to operations in the matrix space, which are in turn reduced to simple operations involving the corresponding symbols. 
For this reason, the reader should not be surprised to find out that almost the entire discussion will be focussed on the symbols of the matrices associated with the noninteracting operators.

A useful property is that the commutator of two quadratic fermionic operators is still quadratic with symbol equal to the commutator of the symbols associated with the two operators. Therefore, the symbol $\mathcal Q^{(n)}(k)$ of a generic (quadratic) $n$-site shift invariant conservation law must commute with the symbol of the Hamiltonian $\mathcal H^{(n)}(k)$. If $\mathcal H^{(n)}(k)$ has $2n$ distinct eigenvalues, $\mathcal Q^{(n)}(k)$ is inevitably a function  $F_{\mathcal Q}^{(n)}$ of $\mathcal H^{(n)}(k)$:
\be\label{eq:funH}
\mathcal Q^{(n)}(k)=F_{\mathcal Q}^{(n)}[\mathcal H^{(n)}(k)]\, .
\ee
For a complete set of symbols commuting with $\mathcal H^{(n)}(k)$ we can choose $F_{\mathcal Q}^{(n)}$ to be a polynomial of order $2n-1$. 
As shown below, we can in fact exploit the one-site shift invariance of the Hamiltonian to keep the linearity in the matrix elements. 

\paragraph{One-site representation.}
In the one-site representation, if $\mathcal H^{(1)}(k)$ is not proportional to the identity (this is a quite unusual situation,  realized \emph{e.g.} by a Hamiltonian with nothing but a Dzyaloshinskii-Moriya interaction), a complete set of one-site shift invariant conservation laws is associated with symbols of the form
\be
\mathcal Q^{(1)}(k)=\lambda_0^{(o)}(k)\mathrm I_2+\lambda_1^{(e)}(k)\mathcal H^{(1)}(k)\, ,
\ee
where $\mathrm I_2$ is the 2-by-2 identity and the superscripts $(e)$ or $(o)$ mean that the (real) function is even or odd respectively (in order to satisfy conditions~\eref{eq:conditions}). 
The local properties of the conservation law can be easily inferred by the fact that the spin representation is directly related to the Fourier transform of the symbol \eref{eq:anti}.
Quasi-locality (we restrict to exponentially localized operators) is therefore assured if $\lambda_i(k)$ are smooth $2\pi$-periodic functions of $k$; locality requires also the functions to have a finite number of nonzero Fourier coefficients\footnote{Here we are using that the Hamiltonian is short-range and hence $\mathcal H^{(n)}(k)$ has only a finite number of Fourier coefficients different from zero.}. A convenient choice that, for any finite chain, produces a complete set of quantum numbers is~\cite{FE_13a} 
\be
\lambda_0^{(o)}(k)\propto\sin(\ell k)\, ,\qquad \lambda_1^{(e)}(k)\propto \cos((\ell-1) k)\, ,\qquad \ell\geq 1\, .
\ee
\paragraph{Two-site representation.}
In the two-site representation the symbol of a one-site shift invariant conservation law reads as
\be\label{eq:CL}
\mathcal Q^{(2)}(k)=\frac{1+\tau_1^x e^{i\frac{k}{2}\tau_1^z}}{2}\mathcal Q^{(1)}(k/2)+\frac{1-\tau_1^x e^{i\frac{k}{2}\tau_1^z}}{2}\mathcal Q^{(1)}(k/2+\pi)\, ,
\ee
where $\mathcal Q^{(1)}(k)$ is its one-site representation. In the absence of degeneracy, only matrices of the form \eref{eq:CL} can commute with $\mathcal H^{(2)}(k)$: there are no additional two-site shift invariant conservation laws.

If  instead $\mathcal H^{(2)}(k)$ is degenerate (which means $\varepsilon(k)=\varepsilon(k+\pi)$, \emph{cf}. \eref{eq:from1to2}), there are matrices commuting with $\mathcal H^{(2)}(k)$ that cannot be written as in \eref{eq:CL}, having nonzero matrix elements between the two sectors identified by the eigenvalues of $\tau_1^x e^{i\frac{k}{2}\tau_1^z}$.

As a matter of fact, degeneracy restricted to isolated momenta is generally not sufficient to produce additional \emph{local} conservation laws, as one can understand by considering that localization in momentum space implies delocalization in real space. 
If instead the dispersion relation satisfies $\varepsilon(k)=\varepsilon(k+\pi)$ for any $k$, the previous argument does not apply and we can find an extra family of local charges with symbol different from \eref{eq:CL}.

\subsection{Example: quantum XY model without magnetic field}
Let us consider the model \eref{eq:HXY0}. In the one-site representation the Hamiltonian has the symbol 
\be
\mathcal H^{(1)}(k)=-J \cos k\ \tau^y+J \gamma\sin k\ \tau^x\, ,
\ee
while in the two-site representation the symbol reads as ($\tau^\alpha_{1,2}$ are Pauli matrices acting on different spaces and $\tau_1^\alpha\tau_2^\beta\equiv \tau^\alpha\otimes\tau^\beta$)
\be
\mathcal H^{(2)}(k)=-\varepsilon_0(k)\tau_1^x e^{i\frac{k}{2}\tau_1^z}\tau_2^y e^{i\theta_0(k)\tau_2^z}
\ee
with
\be\label{eq:eps0}
\fl \quad\varepsilon_0(k)=J \sqrt{\cos^2(k/2)+\gamma^2\sin^2(k/2)}\, ,\qquad e^{i\theta_0(k)}=\frac{\cos(k/2)+i \gamma \sin(k/2)}{\sqrt{\cos^2(k/2)+\gamma^2\sin^2(k/2)}}\, .
\ee
Besides the conservation laws with symbols that commute with $\tau_1^x e^{i\frac{k}{2}\tau_1^z}$, which have the form~\eref{eq:CL}, there is a further class of charges with symbols that instead anticommute with $\tau_1^x e^{i\frac{k}{2}\tau_1^z}$ and have therefore the form
\be
\fl\qquad \tau_1^ye^{i\frac{k}{2}\tau_1^z}(A^{(e)}_{k} \varepsilon_0(k)\tau_2^x e^{i\theta_0(k)\tau_2^z}+B^{(e)}_{k/2} \tau_2^z)+ \tau_1^z(C^{(o)}_{k/2} \varepsilon_0(k)\tau_2^x e^{i\theta_0(k)\tau_2^z}+D^{(o)}_k \tau_2^z)\, .
\ee
The quasi-locality condition enforces $A_k$, $B_k$, $C_k$, and $D_k$ to be smooth $2\pi$-periodic functions of $k$ as well as $B_{k+\pi}=-B_k$ and $C_{k+\pi}=-C_k$. Locality requires instead the functions to have finite numbers of nonzero Fourier coefficients.

Finally, we can choose the following symbols (associated with local charges):
\begin{eqnarray}\label{eq:Q}
Q^{+\rm (e)}_\ell(k)=\cos((\ell-1) k) \varepsilon_0(k) \tau_1^ye^{i\frac{k}{2}\tau_1^z}\tau_2^x e^{i\theta_0(k)\tau_2^z}\nn
Q^{+\rm (o)}_\ell(k)=\cos((\ell-1/2) k)\tau_1^ye^{i\frac{k}{2}\tau_1^z}\tau_2^z\nn
Q^{-\rm (e)}_\ell(k)=\sin(\ell k)\tau_1^z\tau_2^z\nn
Q^{-\rm (o)}_\ell(k)=\sin((\ell-1/2) k)\varepsilon_0(k)\tau_1^z\tau_2^x e^{i\theta_0(k)\tau_2^z}\, .
\end{eqnarray}
For the sake of completeness we also write (the two-site representation of) the symbols of the one-site shift invariant local conservation laws
\begin{eqnarray}\label{eq:I}
I^{+\rm (e)}_\ell(k)=\cos((\ell-1)k)\varepsilon_0(k)\tau_1^xe^{i\frac{k}{2}\tau_1^z}\tau_2^y e^{i\theta_0(k)\tau_2^z}\nn
I^{+\rm (o)}_\ell(k)=\cos((\ell-1/2) k) \varepsilon_0(k)\tau_2^y e^{i\theta_0(k)\tau_2^z}\nn
I^{-\rm (e)}_\ell(k)=\sin(\ell k)\mathrm I\nn
I^{-\rm (o)}_\ell(k)=\sin((\ell-1/2) k)\tau_1^xe^{i\frac{k}{2}\tau_1^z}\, .
\end{eqnarray}
Notice that  in the one-site representation the charges  \eref{eq:I} are generally recast in two classes, as in \eref{eq:claws}.  Symbols \eref{eq:Q} and \eref{eq:I} form a complete set of independent matrices commuting with $\mathcal H^{(2)}(k)$ and producing local conservation laws\footnote{
Since the 2-by-2 identity commutes with 4 independent matrices (which span the entire space) and the symbol of the Hamiltonian has two double degenerate eigenvalues, we can not find more than 8 linearly independent matrices commuting with $\mathcal H^{(2)}(k)$.}. These are not in involution, as in the standard case, but satisfy the following algebra (all the other commutators vanish):
\begin{eqnarray}
\fl i [Q^{+\rm (e)}_\ell,Q^{+\rm (o)}_n]&=I^{+\rm (o)}_{\ell-n}+I^{+\rm (o)}_{\ell+n-1}\nn
\fl i [Q^{-\rm (e)}_\ell,Q^{-\rm (o)}_n]&=-I^{+\rm (o)}_{\ell-n+1}+I^{+\rm (o)}_{\ell+n}\nn
\fl i [Q^{+\rm (e)}_\ell,Q^{-\rm (o)}_n]&=p (I^{-\rm (o)}_{\ell-n}-I^{-\rm (o)}_{\ell+n-1})+\frac{1-p}{2}(I^{-\rm (o)}_{\ell-n+1}+I^{-\rm (o)}_{\ell-n-1}-I^{-\rm (o)}_{\ell+n}-I^{-\rm (o)}_{\ell+n-2})\nn
\fl i [Q^{+\rm (o)}_\ell,Q^{-\rm (e)}_n]&=I^{-\rm (o)}_{\ell-n}-I^{-\rm (o)}_{\ell+n}\nn
\fl i [Q^{+\rm (e)}_\ell,I^{+\rm (o)}_n]&=-p(Q^{+\rm (o)}_{\ell-n}+Q^{+\rm (o)}_{\ell+n-1})-\frac{1-p}{2}(Q^{+\rm (o)}_{\ell-n+1}+Q^{+\rm (o)}_{\ell-n-1}+Q^{+\rm (o)}_{\ell+n}+Q^{+\rm (o)}_{\ell+n-2})\nn
\fl i [Q^{-\rm (e)}_\ell,I^{-\rm (o)}_n]&=-Q^{+\rm (o)}_{\ell-n+1}+Q^{+\rm (o)}_{\ell+n}\nn
\fl i [Q^{+\rm (e)}_\ell,I^{-\rm (o)}_n]&=-Q^{-\rm (o)}_{\ell-n}+Q^{-\rm (o)}_{\ell+n-1}\nn
\fl i [Q^{-\rm (e)}_\ell,I^{+\rm (o)}_n]&=Q^{-\rm (o)}_{\ell-n+1}+Q^{-\rm (o)}_{\ell+n}\nn
\fl i [Q^{+\rm (o)}_\ell,I^{+\rm (o)}_n]&=Q^{+\rm (e)}_{\ell-n+1}+Q^{+\rm (e)}_{\ell+n}\nn
\fl i [Q^{-\rm (o)}_\ell,I^{-\rm (o)}_n]&=-Q^{+\rm (e)}_{\ell-n+1}+Q^{+\rm (e)}_{\ell+n}\nn
\fl i [Q^{-\rm (o)}_\ell,I^{+\rm (o)}_n]&=-p(Q^{-\rm (e)}_{\ell-n}+Q^{-\rm (e)}_{\ell+n-1})-\frac{1-p}{2}(Q^{-\rm (e)}_{\ell-n+1}+Q^{-\rm (e)}_{\ell-n-1}+Q^{-\rm (e)}_{\ell+n}+Q^{-\rm (e)}_{\ell+n-2})\nn
\fl i [Q^{+\rm (o)}_\ell,I^{-\rm (o)}_n]&=-Q^{-\rm (e)}_{\ell-n}+Q^{-\rm (e)}_{\ell+n-1}\, ,
\end{eqnarray}
where $p\equiv (1+\gamma^2)/2$.
Notice that the Hamiltonian is $H=I_1^{(e)}$.
The two classes of conservation laws $I_n^{+(e)}$ and $I_n^{-(e)}$ commute with the entire set, therefore our construction holds true even if we break  reflection symmetry ($I_n^{-(e)}$ are not reflection symmetric) \emph{e.g.} adding a Dzyaloshinskii-Moriya-like interaction 
\be
\frac{1}{4}\sum_\ell \sigma_\ell^x\sigma_{\ell+1}^z\sigma_{\ell+2}^y-\sigma_\ell^y\sigma_{\ell+1}^z\sigma_{\ell+2}^x\, ,
\ee 
which preserves $\varepsilon(k)=\varepsilon(k+\pi)$.

Because of \eref{eq:funH}, given a (finite) linear combination of the local conservation laws with nondegenerate symbol, the other local charges in involution can be obtained by computing the powers of the symbol, multiplying then the result by an oscillatory term that transforms according to \eref{eq:conditions}. 
From a physical point of view this means that any breaking of superintegrability by a local charge $Q$ (\emph{i.e.} perturbing the Hamiltonian with $Q$) ``selects'' a maximal set of local conservation laws that commute between each other (see also \Fref{fig:drawing}). 

The discussion is generalized straightforwardly to higher order representations, which provide further independent local conservation laws if the dispersion relation has exceptional symmetries like $\varepsilon(k)=\varepsilon(k+\frac{2\pi}{n})$, with $n$ integer.

\paragraph{Remark.}
We point out that there is a close similarity between the classification of models as integrable or generic and our classification of noninteracting models as superintegrable or integrable. 

The Hamiltonian of a generic model has a typical nondegenerate spectrum, although degeneracy is not forbidden; in contrast, the Hamiltonian of an integrable model has energy level spacing with a typical Poisson distribution, which signals a highly degenerate spectrum (in the thermodynamic limit).

In our case, the symbol of a generic noninteracting Hamiltonian is nondegenerate, although there may be degeneracy for some isolated momenta; in contrast, the symbol of a `superintegrable' noninteracting Hamiltonian is degenerate for a measurable set of momenta (in the thermodynamic limit).

\section{Time evolution close to superintegrability points}\label{s:time}
The analysis of the previous section reveals that the non-equilibrium time evolution of a state that breaks translation invariance could be subjected to `unusual' constraints. Understanding this problem is quite important, especially in relation to the growing interest in the aspects of the initial state that have an impact on the stationary behavior of observables~\cite{T-HS_13,HR-11, ZSS-12}.  

In addition, because of its simplicity, the XX model (in any equivalent formulation: hard-core bosons, etc.) is often included in the bunch of models analyzed to gain some general insights into the quench problem. However, the very XX model (\eref{eq:HXY0} with $\gamma=0$) has one of the most ``dangerous'' Hamiltonians, having an infinite number of additional local charges that break translation invariance. 

Recognizing the peculiarities of a model is an important but only preliminary step, which prepares to the more ambitious goal of understanding the consequences. In this and in the next section we present a first effect of being close to a superintegrability point: \emph{pre-relaxation}.  

\subsection{The model}
\begin{figure}[htbp]
\begin{center}
\includegraphics[width=0.6\textwidth]{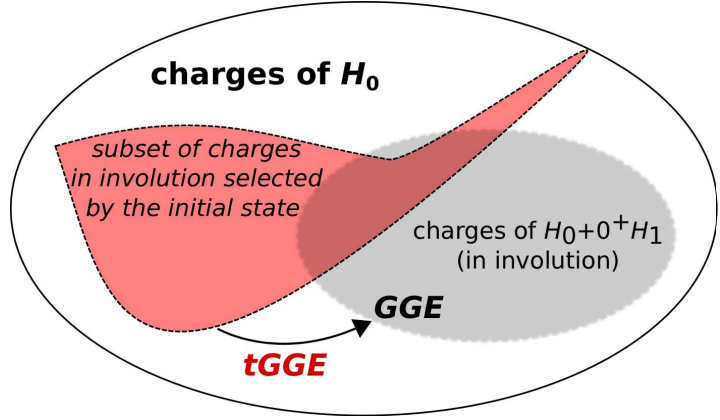}
\caption{Pictorial representation of quantum quenches in models with extra families of local conservation laws. The unperturbed Hamiltonian $H_0$ has a number of local charges (the largest set in the picture) larger than a maximal set of local conservation laws in involution. The local conservation laws are reduced to a subset of charges in involution by an infinitesimal perturbation $H_1$. The generalized Gibbs ensemble associated with $H_0$ can be written in terms of a set of (quasi-)local charges in involution (red set), which is generally different from the set selected by the perturbation (grey set). A time-dependent generalized Gibbs ensemble (tGGE) describes the evolution from the former to the latter set.}
\label{fig:drawing}
\end{center}
\end{figure}

Let us consider the time evolution under the Hamiltonian of the XY model with a small magnetic field $h$
\be\label{eq:HXY}
H(\gamma,h)=J\sum_\ell \Bigl[\frac{1+\gamma}{4} \sigma_\ell^x\sigma_{\ell+1}^x+\frac{1-\gamma}{4} \sigma_\ell^y\sigma_{\ell+1}^y+\frac{h}{2}\sigma_\ell^z\Bigr]\, .
\ee
For the rest of the paper we assume $J>0$.
The symmetry that produces non-translation-invariant local charges is spoiled by the magnetic field $h$.
We are going to show that if the initial state breaks one-site shift invariance then:
\begin{enumerate}
\item The reduced density matrix of $\ell\ll h^{-1}$ neightboring spins approaches a stationary value in the limit
\be\label{eq:tw}
\ell\ll J t\ll h^{-1}\, ;
\ee
\item At later times one-site shift invariance is restored;
\item The relaxation following the pre-relaxation regime can be approximately described in terms of a ``time-dependent GGE'' of the form \eref{eq:GGEimp} that can be written in terms of the local conservation laws of the unperturbed Hamiltonian $H(\gamma,0)$. 
\end{enumerate}
\Fref{fig:drawing} depicts the process that we are going to describe. 

\subsection{Exact solution}
In order to have full control of the errors arising from the approximations, we first present the exact solution of the quench problem starting from a two-site shift invariant state. 
We write the Hamiltonian in the 4-by-4 block-circulant form \eref{eq:H4}. 
The two-site representation of the correlation matrix $\Gamma$ is defined as
\be
\Gamma_{\ell n}=\delta_{\ell n}\mathrm I_4-\Braket{\left(\begin{array}{c}a_{2\ell-1}^x\\a^{y}_{2\ell-1}\\a_{2\ell}^x\\a_{2\ell}^y
\end{array}\right)\left(\begin{array}{cccc}a_{2n-1}^x&a^{y}_{2n-1}&a_{2n}^x&a_{2n}^y
\end{array}\right)}\, .
\ee
where $\mathrm I_4$ is the 4-by-4 identity. 
The time dependent correlation matrix is then given by
\begin{eqnarray}\label{eq:cmt}
\fl\qquad\qquad&\Gamma_{\ell n}(t)&=\frac{2}{L}\sum_{k} e^{-i(n-\ell)k}\Gamma(k;t)\nn
\fl\qquad\qquad&&=[e^{-i \mathcal H t}\Gamma(0)e^{i \mathcal H t}]_{\ell n}=\frac{2}{L}\sum_{k} e^{-i(n-\ell)k} e^{-i \mathcal H^{(2)}(k)t}\Gamma(k;0)e^{i \mathcal H^{(2)}(k) t}\, ,
\end{eqnarray}
where $\Gamma(0)$ is the initial correlation matrix and $\Gamma(k;0)$ its symbol.

The symbol of the Hamiltonian~\eref{eq:HXY} reads as
\be
\mathcal H^{(2)}(k)=J h \tau_2^y-\varepsilon_0(k)\tau_1^x e^{i\frac{k}{2}\tau_1^z}\tau_2^y e^{i\theta_0(k)\tau_2^z}\, ,
\ee
with $\varepsilon_0(k)$ and $e^{i\theta_0(k)}$ of \eref{eq:eps0}, and generates the time evolution matrix
\be\label{eq:teo}
e^{-i \mathcal H^{(2)}(k) t}=\sum_{s=\pm 1}\Pi_1^s(k) e^{-i \theta_{s}(k)\tau_2^z/2}e^{is \varepsilon_s(k)\tau_2^y t}e^{i \theta_{s}(k)\tau_2^z/2}
\ee
where $\Pi_1^\pm(k)$ are projectors
\be
\Pi_1^s(k)=\frac{1+ s\tau_1^x e^{i\frac{k}{2}\tau_1^z}}{2} 
\ee
and
\begin{eqnarray}
\varepsilon_s(k)=J \sqrt{(s h-\cos(k/2))^2+\gamma^2\sin^2(k/2)}\, ,\nn
 e^{i\theta_s(k)}=\frac{-s h+\cos(k/2)+i \gamma \sin(k/2)}{\sqrt{(s h-\cos(k/2))^2+\gamma^2\sin^2(k/2)}}\, .
\end{eqnarray}
Substituting \eref{eq:teo} into \eref{eq:cmt} gives the time dependent correlation matrix. 

We are interested in the relaxation properties of subsystems, thus we must identify the terms of $\Gamma(k;t)$ that give nonzero contribution in the limit $\ell\ll J t$, where $\ell$ is the subsystem length. This can be done as follows. 

For the Wick theorem, the expectation value of any operator with an even number of fermions can be written in terms of the correlation matrix of the subsystem in which the operator acts (nontrivially).  
The correlation matrix of a subsystem of $\ell$ contiguous spins is a square submatrix on the diagonal of the total correlation matrix with $2\ell$ rows (\emph{i.e.} it is a $(2n)$-by-$(2n)$ block-Toeplitz matrix with $(\ell/n)^2$ block-elements, where $n$ is the index of the representation).
The (block-)elements are the Fourier coefficients of the symbol restricted to the smallest frequencies, so terms in $\Gamma(k;t)$  with rapidly oscillating phases give a negligible contribution in the limit in which the oscillation frequency is large compared to $\ell$. This allows us to neglect the terms proportional to $e^{i\varepsilon_s(k)t}$, provided that the time is sufficiently large.
We finally end up with a stationary term ($\mathrm{tr}_{\{2\}}$ is the trace over the space in which $\tau_2^\alpha$ act)
\be
\fl\qquad\Gamma_{\rm GGE}(k)=\sum_{s=\pm1}\Pi_1^s(k)\Bigl(\frac{1}{2}\tr{\{2\}}{\Gamma(k,0)}+\frac{1}{2}\tr{\{2\}}{\Gamma(k,0)\tau_2^ye^{i\theta_s(k)\tau_2^z}}\tau_2^ye^{i\theta_s(k)\tau_2^z}\Bigr)\Pi_1^s(k)
\ee
and a time dependent one
\begin{eqnarray}
\fl\quad\Gamma_{\rm sym}(k;t)=\sum_{s=\pm1}\Pi_1^s(k) e^{- i\frac{\theta_s(k)}{2}\tau_2^z}\Bigl(\frac{1}{2}\tr{\{2\}}{\Gamma(k,0)\tau_2^z e^{i\frac{\theta_s(k)-\theta_{-s}(k)}{2}\tau_2^z}}\tau_2^z\nn
+\frac{1}{2}\tr{\{2\}}{\Gamma(k,0)\tau_2^xe^{i\frac{\theta_s(k)+\theta_{-s}(k)}{2}\tau_2^z}}\tau_2^x\Bigr)e^{i s(\varepsilon_{-s}(k)-\varepsilon_{s}(k))t\tau_2^y}e^{i\frac{\theta_s(k)}{2}\tau_2^z}\Pi_1^{-s}(k)\, ,
\end{eqnarray}
which can not be neglected because the time is multiplied by a term that approaches zero as our small parameter $h\rightarrow 0$. At the leading order in $h$ we indeed have
\be
\fl\quad e^{i\theta_s(k)\tau_2^z}=e^{i\theta_0(k)\tau_2^z}+O(h)\, ,\qquad\lim_{h\rightarrow 0}s \frac{\varepsilon_{-s}(k)-\varepsilon_{s}(k)}{h}=2 \mathcal E(k) \equiv 2 J \cos\theta_0(k)\, ,
\ee
and hence
\begin{eqnarray}\label{eq:GammaGGE}
\fl\quad\Gamma_{\rm GGE}(k)\sim \sum_{s=\pm1}\Pi_1^s(k)\Bigl(\frac{1}{2}\tr{\{2\}}{\Gamma(k;0)}+\frac{1}{2}\tr{\{2\}}{\Gamma(k;0)\tau_2^ye^{i\theta_0(k)\tau_2^z}}\tau_2^ye^{i\theta_0(k)\tau_2^z}\Bigr)\Pi_1^s(k)\nn
\fl\quad\Gamma_{\rm sym}(k;t)\sim \sum_{s=\pm1}\Pi_1^s(k)e^{- i\frac{\theta_0(k)}{2}\tau_2^z}\Bigl(\frac{1}{2}\tr{\{2\}}{\Gamma(k;0)\tau_2^z }\tau_2^z+
\frac{1}{2}\tr{\{2\}}{\Gamma(k;0)\tau_2^xe^{i\theta_0(k)\tau_2^z}}\tau_2^x\Bigr)\nn
e^{2i \mathcal E(k)(h t)\tau_2^y}e^{i\frac{\theta_0(k)}{2}\tau_2^z}\Pi_1^{-s}(k)\, .
\end{eqnarray}
These expressions have a double meaning: on the one hand, in the limit $\ell\ll J t\ll h^{-1}$ the correlation matrix approaches a stationary value
\be\label{eq:p-GGE}
\Gamma(k;t)\rightarrow \Gamma_{\rm GGE}(k)+\Gamma_{\rm sym}(k;0)\, ,
\ee
which would have been the GGE correlation matrix for zero magnetic field;
on the other hand,  at larger times  the correlation matrix is well approximated by 
\be
\Gamma(k;t)\sim\Gamma_{\rm GGE}(k)+\Gamma_{\rm sym}(k;t)\, .
\ee
Eventually, the contribution from $\Gamma_{\rm sym}(k;t)$ disappears and we end up with $\Gamma_{\rm GGE}(k)$.
\paragraph{Example.}
Let us consider a quench from the Majumdar-Ghosh dimer product state 
\be\label{eq:MGs}
\ket{\Psi_0}=\frac{\ket{\uparrow\downarrow}-\ket{\downarrow\uparrow}}{\sqrt{2}}\otimes\cdots\otimes \frac{\ket{\uparrow\downarrow}-\ket{\downarrow\uparrow}}{\sqrt{2}}\, ,
\ee
which is the ground state of the (translation invariant) Hamiltonian
\be
H_{\rm MG}=
\frac{J}{4}\sum_\ell  \Bigl(\vec\sigma_\ell\cdot \vec \sigma_{\ell+1}+\frac{1}{2}\vec\sigma_\ell\cdot \vec \sigma_{\ell+2}\Bigr)\, ,
\ee 
but also the ground state of the noninteracting model
\be
H_{\rm XX}^{(e)}=\frac{J}{4}\sum_\ell\sigma_{2\ell-1}^x\sigma_{2\ell}^x+\sigma_{2\ell-1}^y\sigma_{2\ell}^y\, .
\ee
As a consequence, the non-equilibrium evolution under the XY Hamiltonian is noninteracting in the Jordan-Wigner fermions~\eref{eq:JW}, and the state~\eref{eq:MGs} is a two-site shift invariant Slater determinant with a very simple correlation matrix with symbol\footnote{There is a simple relation between a noninteracting Hamiltonian and the correlation matrix of its ground state: $\Gamma=-\mathrm{sgn}[\mathcal H]$ (see \cite{FC:2010} for a more general relation). That is to say, the symbol of the correlation matrix is minus the symbol of the Hamiltonian, with the dispersion relation replaced by $1$.}
\be\label{eq:MGcm}
\Gamma_{\rm MG}(k;0)=\tau_1^x \tau_2^y\, .
\ee
From \eref{eq:GammaGGE} we obtain
\begin{eqnarray}\label{eq:corr0}
\fl\qquad\Gamma_{\rm GGE}(k)\sim \cos\theta_0(k)\cos\Bigl(\frac{k}{2}\Bigr)\tau_1^x e^{i\frac{k}{2}\tau_1^z}\tau_2^y e^{i \theta_{0}(k)\tau_2^z}\nn
\fl\qquad\Gamma_{\rm sym}(k;t)\sim-\sin\theta_0(k)\sin\Bigl(\frac{k}{2}\Bigr)\tau_1^y e^{i\frac{k}{2}\tau_1^z}e^{- i\frac{\theta_0(k)}{2}\tau_2^z}\tau_2^x e^{2i \mathcal E(k)(h t)\tau_2^y}e^{i\frac{\theta_0(k)}{2}\tau_2^z}\, .
\end{eqnarray}
The GGE correlation matrix  $\Gamma_{\rm GGE}(k)$ is one-site shift invariant, but $\Gamma_{\rm sym}(k;0)$ is not: in the pre-relaxation regime ($t\rightarrow\infty$, $h J t\rightarrow 0$) translation invariance is not  restored!
Note however that for this particular initial state the time evolution with the XX Hamiltonian does not exhibit pre-relaxation ($\sin\theta_0(k)=0$ for $\gamma=0$).

\subsection{General formalism}
We now show that the results obtained in the previous section can be understood at a more general level.
For the sake of simplicity we still refer to Hamiltonian \eref{eq:HXY}, however the discussion can be easily generalized to other Hamiltonians with a weak perturbation that breaks superintegrability. 

The first step is to single out the unperturbed Hamiltonian with extra families of local conservation laws, which in our case is $H(\gamma,0)$. Since $H(\gamma,h) =H(\gamma,0)+H(0,h)  $ we have
\be\label{eq:Usplit}
e^{-i H(\gamma,h) t}=\Bigl[{\rm T^\dag}\exp\Bigl(-i\int_0^{h J t}\mathrm d \tau V(\tau)\Bigr)\Bigr]e^{-i H(\gamma,0)t}\, ,
\ee
where ${\rm T^\dag}$ is the anti-time-ordering operator (we indicate with ${\rm T}$ the time-ordering one), 
\be \label{eq:Vfirst}
J V(\tau)= e^{-i \frac{H(\gamma,0)}{J h}\tau}H(0,1)e^{i \frac{H(\gamma,0)}{J h}\tau}
\ee
and we used $H(0,h)=h H(0,1)$. 

Despite \eref{eq:Vfirst} is apparently not local, in the limit of small $h$ it is in fact a quasi-local operator. 
The symbol of $V$ is indeed given by
\be
\fl\quad\mathcal V(k;\tau)=\cos\theta_0(k)\tau_2^y e^{i\theta_0\tau_2^z}+\sin\theta_0(k)\Bigl(e^{\frac{2i \varepsilon_0(k)}{J h}\tau}\frac{\tau_2^x e^{i\theta_0(k)\tau_2^z}-i \tau_1^x e^{i\frac{k}{2}\tau_1^z}\tau_2^z}{2}+{\rm h.c.}\Bigr)\, .
\ee
The first term is associated with the projection of $H(0,1) $ onto the eigenspaces of the unperturbed Hamiltonian. 
The contribution from the remaining terms approaches zero in the limit $h\rightarrow 0$, as the phase $e^{2i \varepsilon_0(k)\tau/(J h)}$ is rapidly oscillating (for the elements close to the main diagonal of $\mathcal V$ we can apply a stationary phase approximation; the far away elements decay exponentially with the range and can be neglected anyway).

In the limit of small $h$ the symbol is therefore well approximated by the stationary term
\be
{\mathcal V}(k;\tau)\sim {\mathcal V}_0(k)=\cos\theta_0(k)\tau_2^y e^{i\theta_0\tau_2^z}\, .
\ee
We notice that in the thermodynamic limit $\mathcal V_0$ has the (block-)elements (it acts like the identity on the space of $\tau_1^\alpha$)
\be\label{eq:Vql}
\fl\qquad\qquad[{\mathcal V}_{0}]_{\ell n}=\frac{\delta_{\ell n}}{1+\gamma^2}\tau_2^y+i\mathrm{sgn}(\ell-n)\frac{2\gamma}{(1+|\gamma|)^2}\Bigl(\frac{|\gamma|-1}{|\gamma|+1}\Bigr)^{|\ell-n|-1}\tau_2^{\mathrm{sgn}(n-\ell)}\, ,
\ee
where $\tau_2^{\pm}=(\tau_2^x\pm i\tau_2^y)/2$.
The corresponding fermionic operator $V_0$ is therefore a quasi-local operator with characteristic range $\bar r=2/|\log{\frac{\gamma+1}{\gamma-1}}|$ (for $\gamma\neq 0$). 

Let us now consider the expectation value of a generic local operator $\mathcal O$
\be\label{eq:temp}
\braket{\Psi_0|e^{i H(\gamma,h)t}\mathcal O e^{-i H(\gamma,h)t}|\Psi_0}=\braket{\Psi_0|e^{i H(\gamma,0)t}\mathcal O_t e^{-i H(\gamma,0)t}|\Psi_0}
\ee
where
\be
\mathcal O_t=\Bigl[{\rm T}\exp\Bigl(i\int_0^{h J t}\mathrm d \tau V(\tau)\Bigr)\Bigr]\mathcal O \Bigl[{\rm T^\dag}\exp\Bigl(-i\int_0^{h J t}\mathrm d \tau  V(\tau)\Bigr)\Bigr]\, .
\ee
By virtue of \eref{eq:Vql}, in the scaling limit of small $h$ and finite $h J t $ the operator $\mathcal O_t$ is presumably exponentially localized; on the other hand the time is large, thus, following \cite{EEF:12}, the time evolving state in \eref{eq:temp} can be approximately replaced by the corresponding GGE
\be
\fl\qquad\braket{\Psi_0|e^{i H(\gamma,0)t}\mathcal O_t e^{-i H(\gamma,0)t}|\Psi_0}\sim \tr{}{\rho_{\rm GGE}^{(\gamma,h=0,\Psi_0)}\mathcal O_t}=\tr{}{\rho_{\rm GGE}^{(\gamma,h=0,\Psi_0)}(t)\mathcal O}\, ,
\ee 
where we defined the mixed state
\be
\fl\qquad\rho_{\rm GGE}^{(\gamma,h=0,\Psi_0)}(t)=\Bigl[{\rm T^\dag}\exp\Bigl(-i\int_0^{h J t}\!\!\!\mathrm d \tau  V(\tau)\Bigr)\Bigr]\rho_{\rm GGE}^{(\gamma,h=0,\Psi_0)}\Bigl[{\rm T}\exp\Bigl(i\int_0^{h J t}\!\!\!\mathrm d \tau  V(\tau)\Bigr)\Bigr]\, ,
\ee
which in the scaling limit of small $h$ and finite $h J t$ reads as
\be\label{eq:UtGGE}
\rho_{\rm GGE}^{(\gamma,h=0,\Psi_0)}(t)\sim e^{-i V_0 h J t} \rho_{\rm GGE}^{(\gamma,h=0,\Psi_0)}e^{i V_0 h J t}\, .
\ee

As a matter of fact, this is exactly the same approximation done in the previous section, indeed one can easily show 
\be
\fl\qquad\qquad\Gamma_{\rm sym}(t,k)\sim e^{-i {\mathcal V}_0(k) h J t}\Gamma_{\rm sym}(0,k)e^{i {\mathcal V}_0(k) h J t}\qquad [\Gamma_{\rm GGE}(k),{\mathcal V}_0(k)]=0\, .
\ee
The advantage of the new perspective is that it can in principle be applied also to interacting models. 

In conclusion, the relaxation following pre-relaxation in the limit $h\ll 1$ can be described by the time-dependent generalized Gibbs ensemble
\be\label{eq:tdGGE}
\frac{1}{Z}\exp\Bigl(-\sum_{j=1}\lambda_j e^{- i  V_0 h J t}Q_j e^{i  V_0 h J t}\Bigr)\, ,
\ee
where $\lambda_j$ are the Lagrange multipliers of the GGE that describes the stationary properties in the pre-relaxation time window.
Notice that this is a formal and not very intuitive representation, being written in terms of the quasi-local charges $ e^{- i  V_0 h J t}Q_j e^{i  V_0 h J t}$; however, the time dependent GGE can be expressed at any time in terms of the local conservation laws \eref{eq:Q} and \eref{eq:I}.

\subsection{Numerical results}
\begin{figure}[htbp]
\begin{center}
\includegraphics[width=0.49\textwidth]{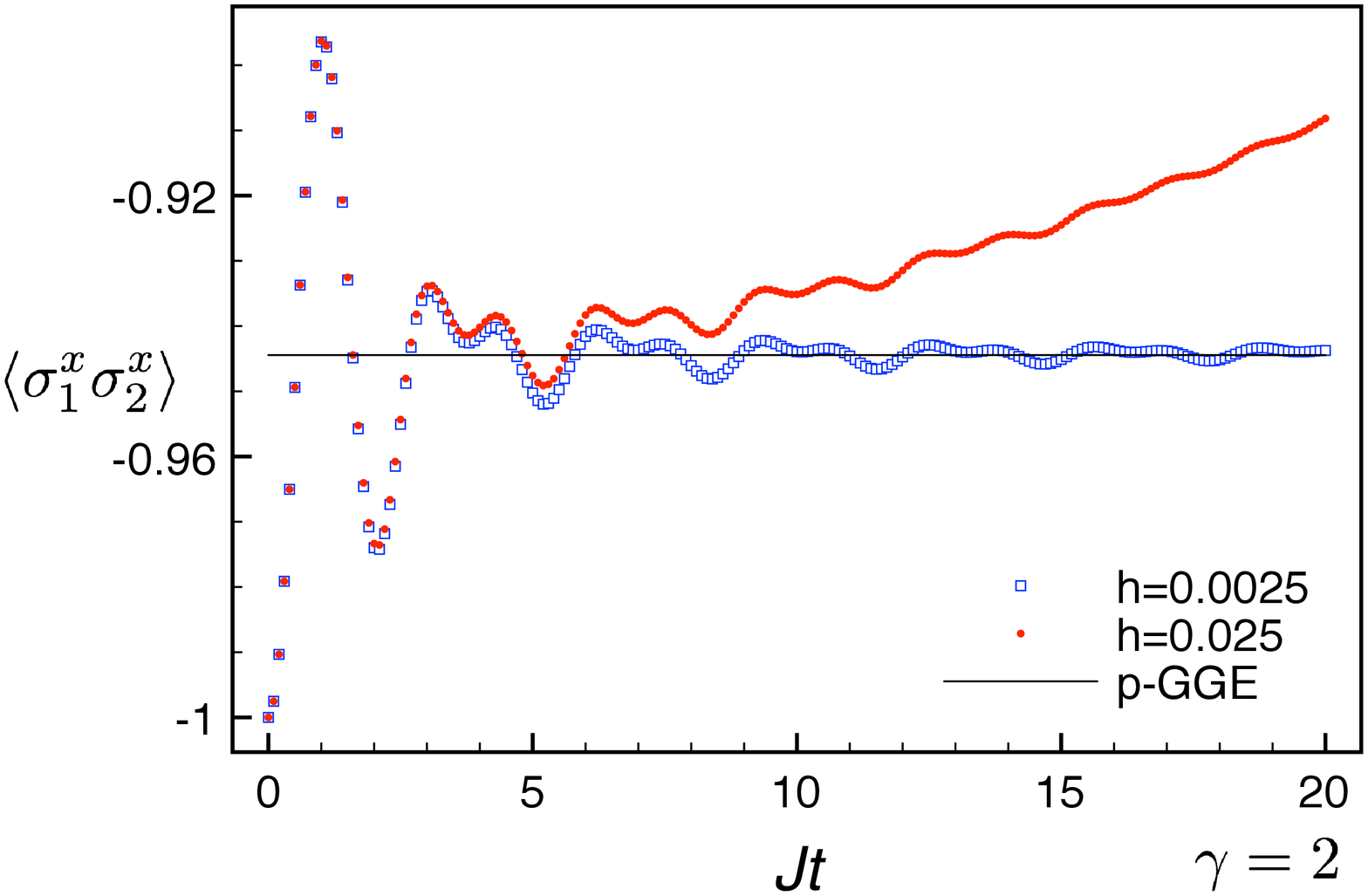}
\includegraphics[width=0.49\textwidth]{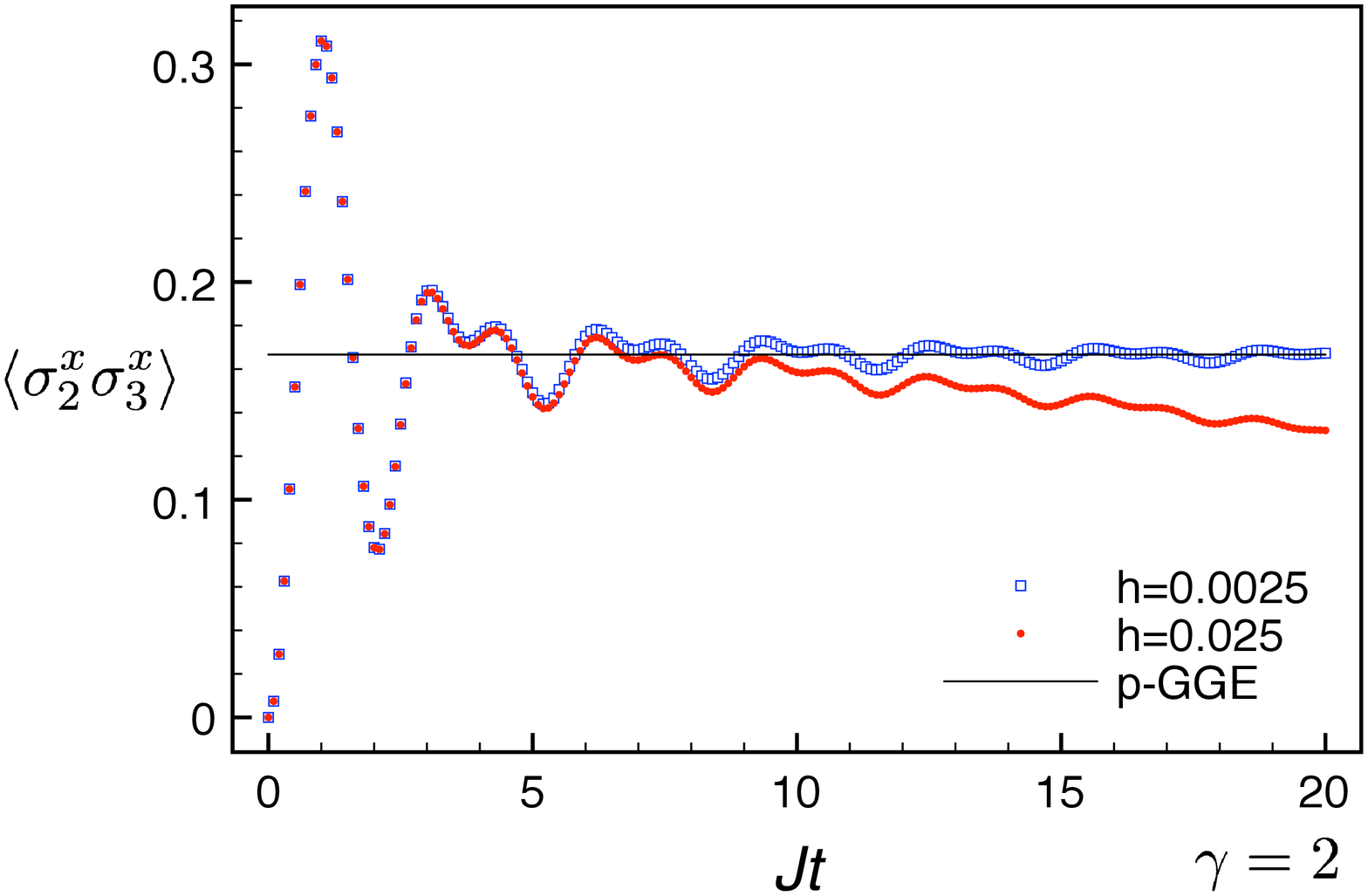}\\
\includegraphics[width=0.49\textwidth]{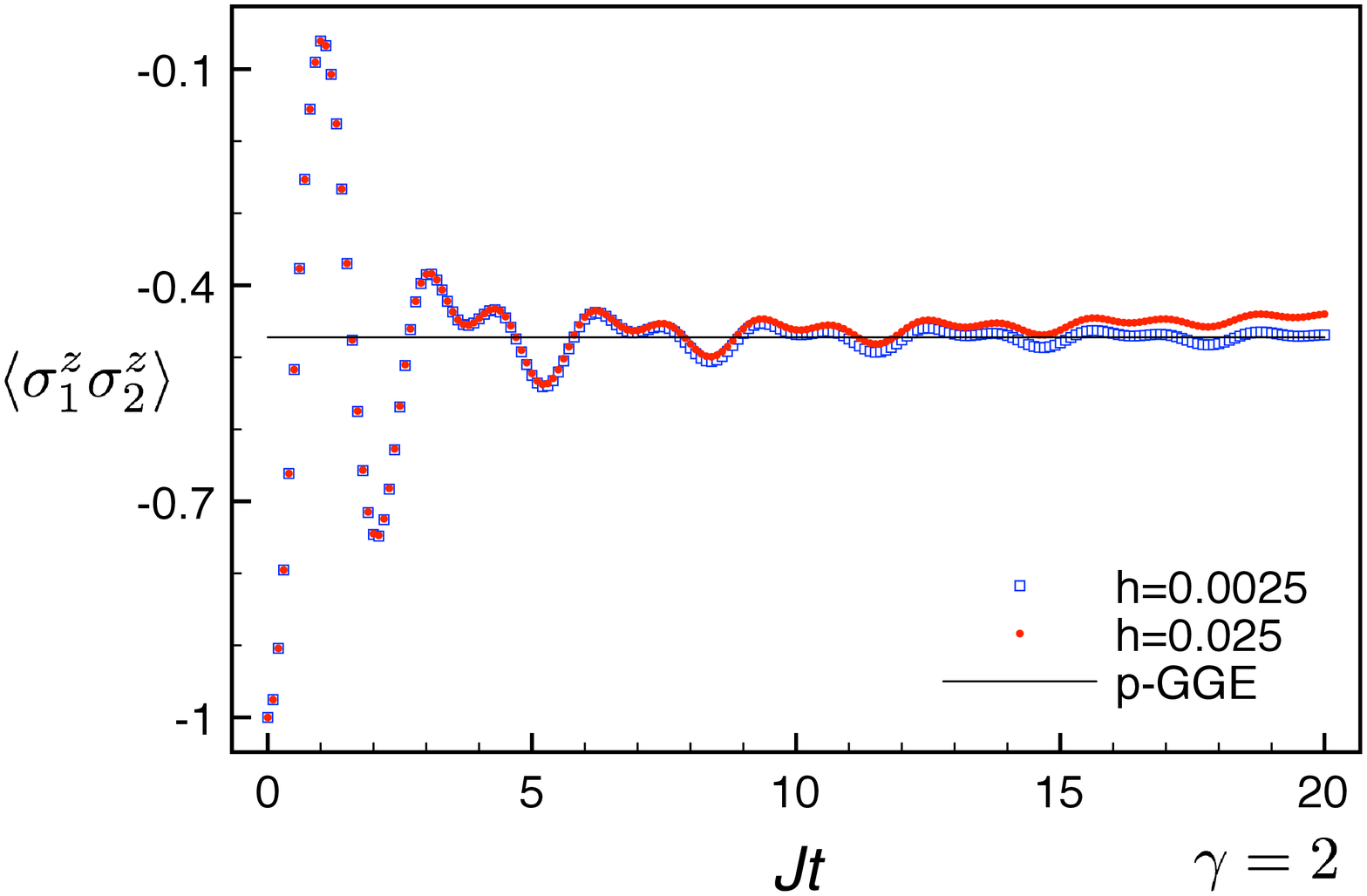}
\includegraphics[width=0.49\textwidth]{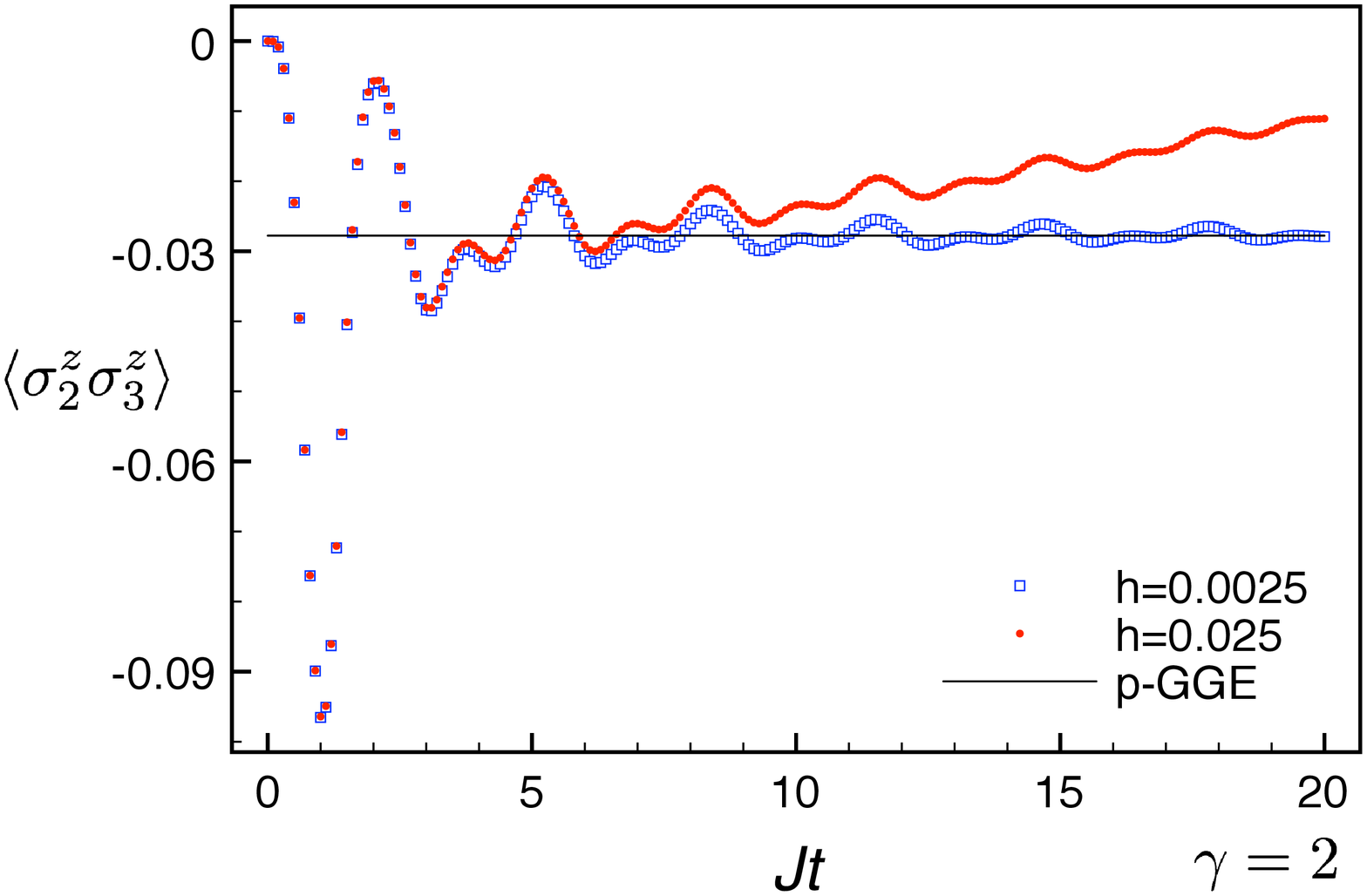}
\caption{Nearest neighbor correlators for two quenches from the Majumdar Ghosh ground state in the quantum XY model with $\gamma=2$ and magnetic field $h=0.0025$ and $h=0.025$. The solid lines (p-GGE) are the GGE values corresponding to the unperturbed model. For the quench with $h=0.0025$ the correlators display a clear pre-relaxation behavior. Notice that the correlators are not translation invariant.}
\label{fig:XYn}
\end{center}
\end{figure}
\begin{figure}[htbp]
\begin{center}
\includegraphics[width=0.49\textwidth]{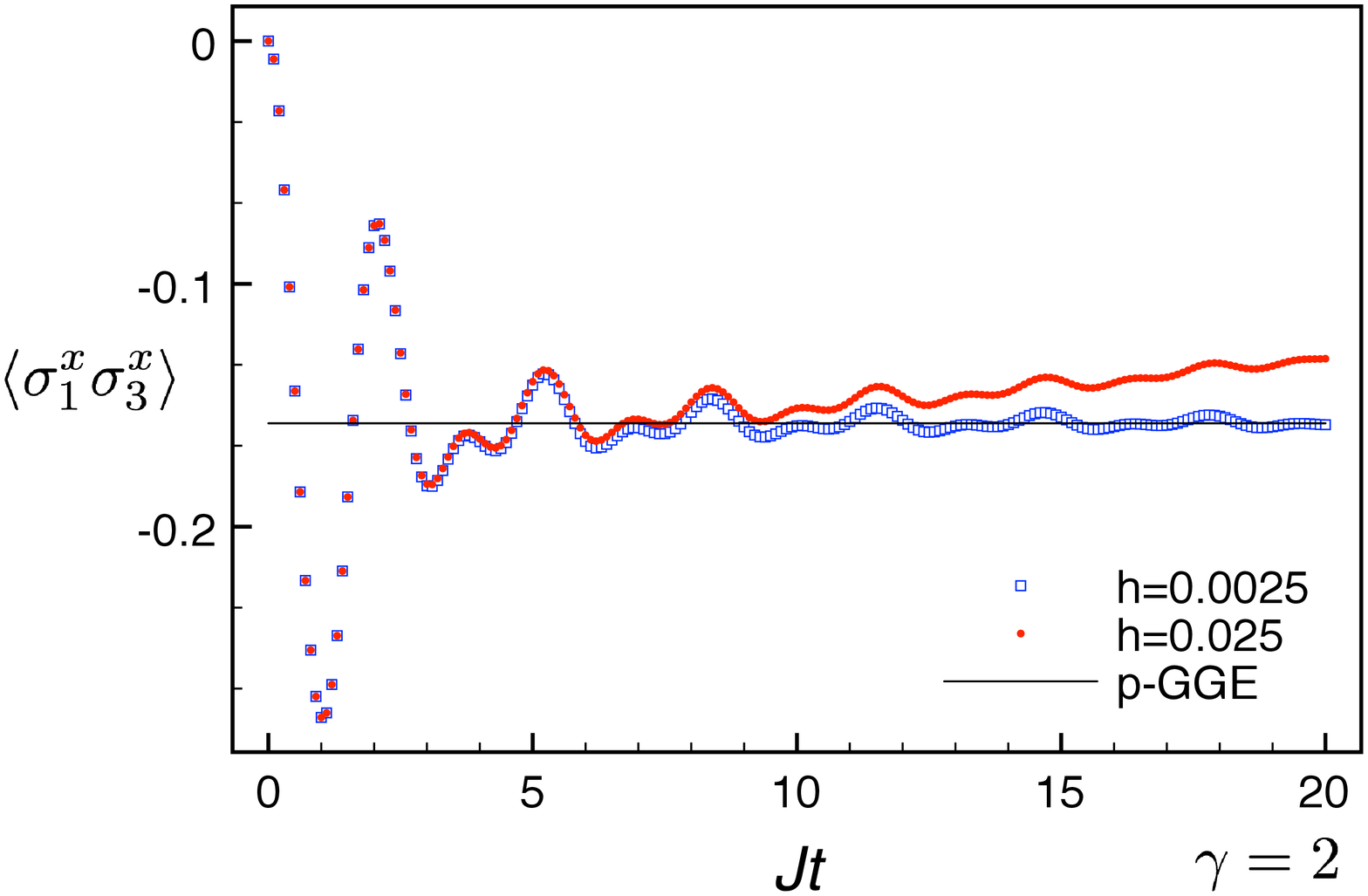}
\includegraphics[width=0.49\textwidth]{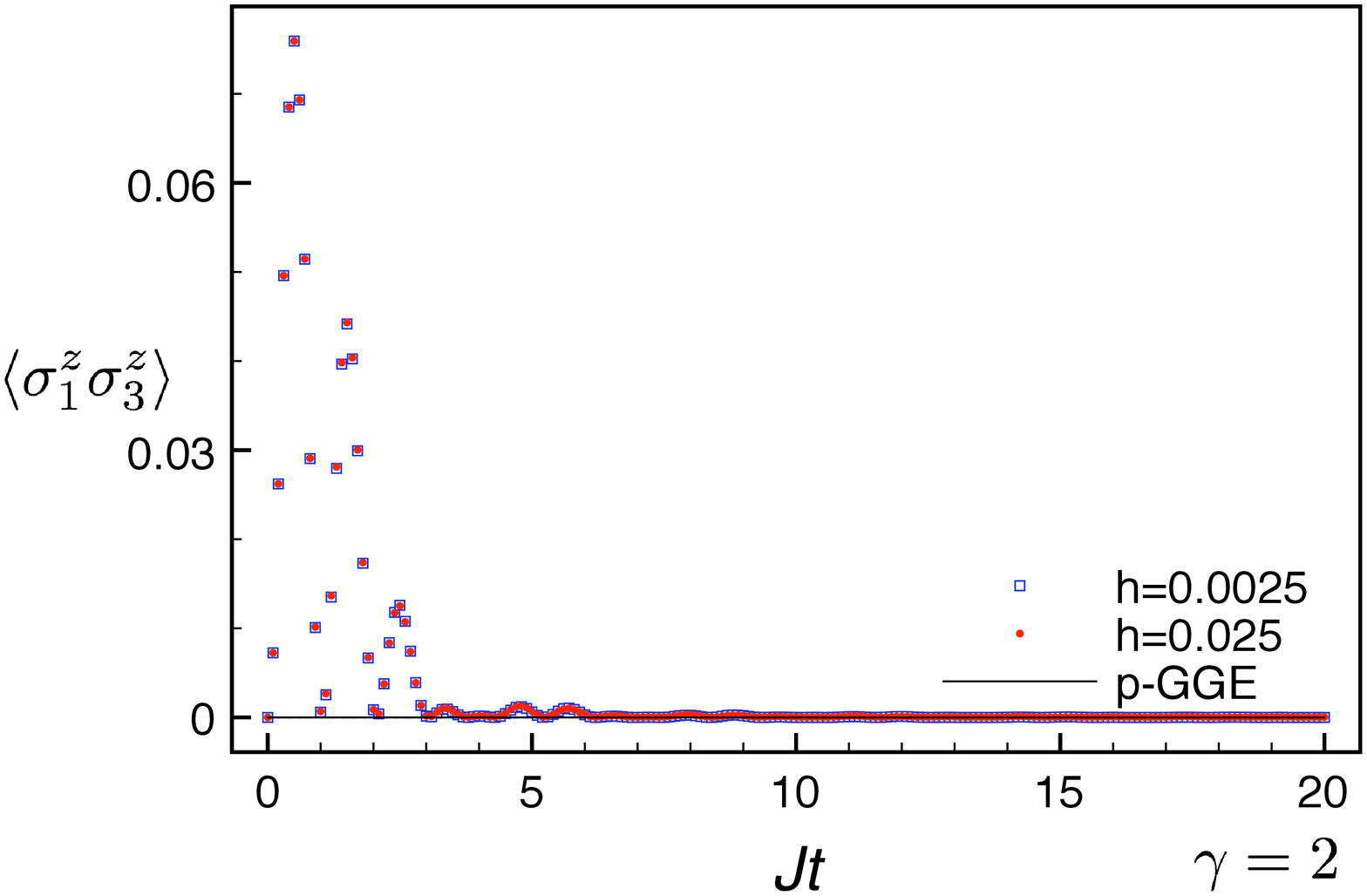}
\caption{Next-nearest neighbor correlators for the quenches of \Fref{fig:XYn} (with the same notations). The correlators are translation invariant by reflection symmetry. The expectation value of $\braket{\sigma_1^z\sigma_3^z}$ in the time dependent GGE is zero at any time.}
\label{fig:XYnn}
\end{center}
\end{figure}

Figures \ref{fig:XYn} and \ref{fig:XYnn} show the time evolution of short-range correlators for two small values of the magnetic field. In the time window considered the correlators of the quench with the smallest $h$ approach the values predicted by the generalized Gibbs ensemble of the unperturbed model~\eref{eq:p-GGE}. In the other quench the discrepancy is visible also at small times (but is $O(h)$!) and the correlators do not experience pre-relaxation. 

\begin{figure}[htbp]
\begin{center}
\includegraphics[width=0.49\textwidth]{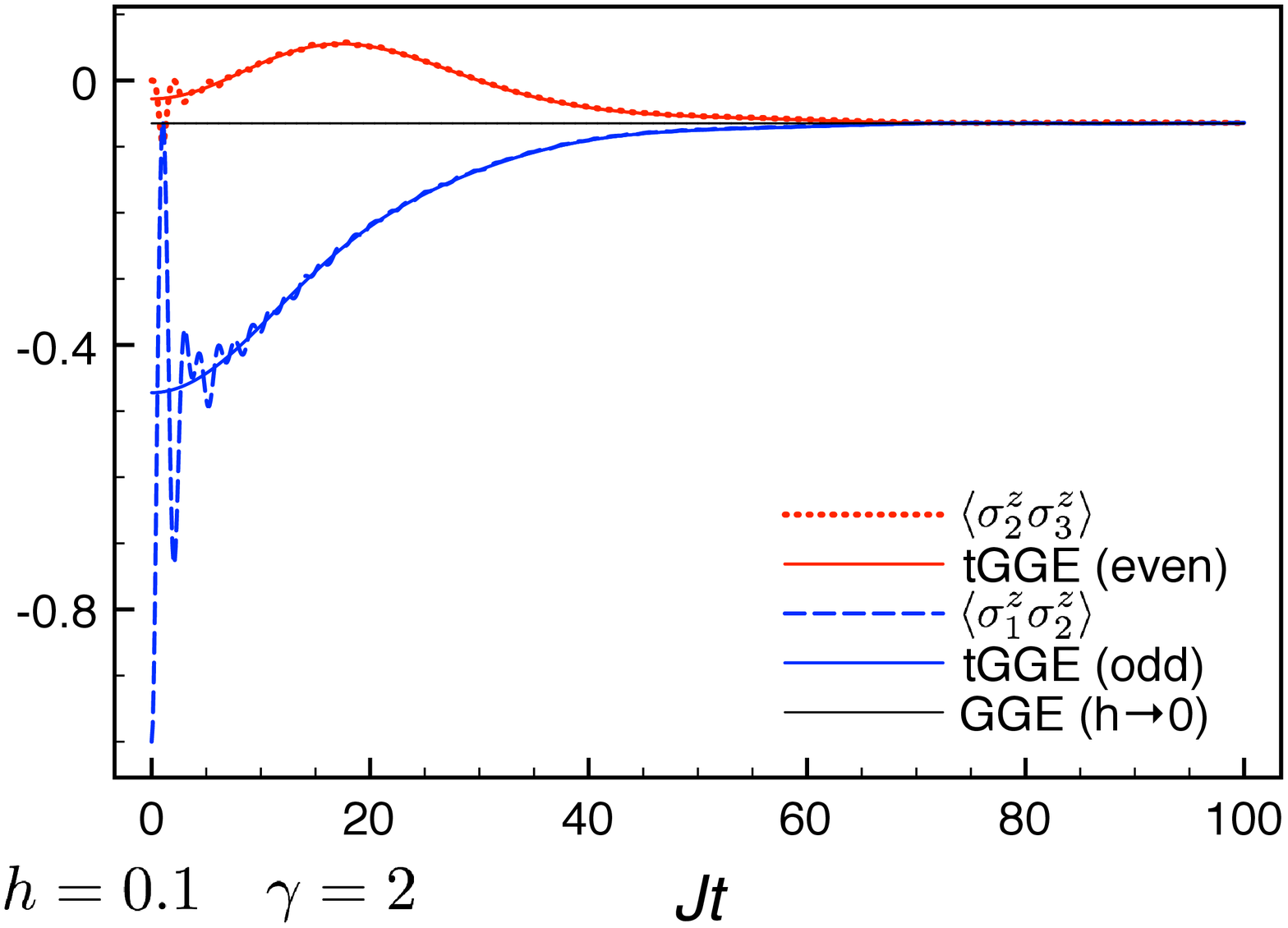}
\includegraphics[width=0.49\textwidth]{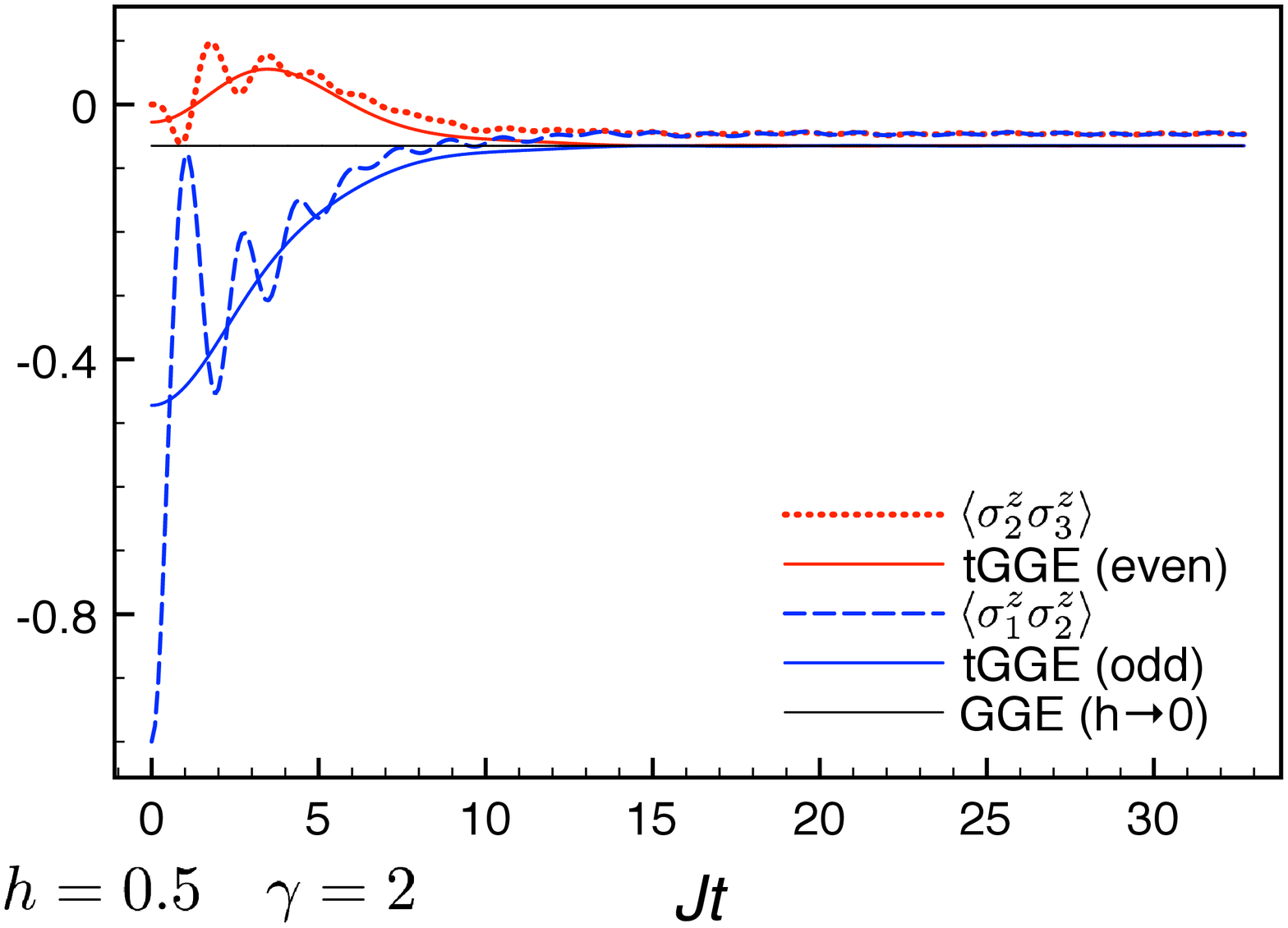}
\caption{Nearest neighbor correlators $\braket{\sigma_\ell^z \sigma_{\ell+1}^z}$ for a quench from the Majumdar Ghosh ground state in the quantum XY model with $\gamma=2$ and magnetic fields $h=0.1$ (left) and $h=0.5$ (right). The solid lines (tGGE) are the values computed in the time-dependent GGE. Left: Despite the magnetic field is not tiny (and hence, there is no pre-relaxation behavior, \emph{cf}. \Fref{fig:XYn}), at sufficiently large times the expectation values agree with the tGGE values. At late times translation invariance is restored and the expectation values are in excellent agreement with the GGE value computed in the limit $h\rightarrow 0$ (black tiny solid line). Right: For a larger magnetic field the agreement is not quantitatively good, however the qualitative behavior is correctly captured.}
\label{fig:XYfull}
\end{center}
\end{figure}

On the other hand, the description in terms of a time-dependent generalized Gibbs ensemble is good even for larger magnetic fields (\Fref{fig:XYfull}; notice the different time scale, $J t\sim h^{-1}$, with respect to Figures \ref{fig:XYn} and \ref{fig:XYnn}). By increasing further the magnetic field (\emph{e.g.} $h=0.5$ in \Fref{fig:XYfull}) the time dependent GGE (which is defined in terms of the conservations laws in the limit $h\rightarrow 0$) provides only a qualitatively good description of the actual time evolution. 
Nevertheless, \Fref{fig:XYfull} clearly shows that \emph{the main process active at large times is the relaxation of quasi-conserved local operators}, which are the local charges of the unperturbed Hamiltonian that can not be obtained as a deformation of the charges of the perturbed one.

\subsection{Recap}
Let us summarize the results obtained in this section.

In the limit of weak perturbation the dynamics following pre-relaxation can be described as follows:
\begin{enumerate}
\item rescale the time by incorporating the time scale introduced by the perturbation (in the case considered $t\rightarrow h t$);
\item replace the initial state with the generalized Gibbs ensemble associated with the unperturbed Hamiltonian $H_0$;
\item time evolve with an effective Hamiltonian $H_{\rm I}$ such that 
\be
e^{-i H_{\rm I} h t}\sim e^{-i (H_0+h H_1) t}e^{i H_0 t} \qquad (h\rightarrow 0,\ t\rightarrow\infty,\ ht\ {\rm finite})\, ,
\ee
where $h H_1$ is the perturbation.
\end{enumerate}
This procedure defines a \emph{time-dependent generalized Gibbs ensemble}. 
An ensemble with time dependent Lagrange multipliers was used in \cite{SK:13} as ansatz to describe the thermalization process in a weakly interacting model. 
In our case, we can clearly identify the conditions under which our construction is nontrivial (\emph{i.e.} a time-dependent GGE makes sense): 
\begin{enumerate}[(a)]
\item the set of local conservation laws of the unperturbed Hamiltonian is larger than the set of the perturbed one;
\item \label{notherm} the set associated with the perturbed Hamiltonian (in the limit of infinitesimal perturbation) does not commute with the entire set of charges of the unperturbed model.
\end{enumerate}
Notice that condition \eref{notherm} comes from the fact that if the perturbation is associated (by \eref{eq:Vfirst}, in the limit of infinitesimal perturbation) with a charge $V_0$ that commutes with all the charges of the unperturbed model, then $V_0$ will also commute with the generalized Gibbs ensemble that describes the pre-relaxation time-window, making \eref{eq:UtGGE} trivially independent of time. 

Because of \eref{notherm}, we do not expect that the same mechanism could contribute to the relaxation process in nonintegrable models close to integrability points: if the local degrees of freedom are going to thermalize, the perturbation will restrict the set of conservation laws to the single Hamiltonian, which commutes with the entire original set of charges by definition. 

However, it is less clear whether a superintegrable model with a perturbation that breaks integrability  might instead experience a similar form of pre-relaxation. In principle, we can identify two pre-relaxation regimes: a first, in which the perturbation can be ignored, and a second, in which the integrability breaking term can be effectively replaced by some integrable perturbation that breaks only superintegrability. 

We notice that this situation is not completely new: \emph{e.g.} \cite{EKMR-13} considered the time evolution under the Hamiltonian of a non-integrable model close to a superintegrable point, even though this aspect was not highlighted.

\section{Quantum quenches close to the Ising limit of the XXZ spin-1/2 chain }\label{s:time1}
In this section we consider the time evolution under the Hamiltonian of the spin-1/2 XXZ Heisenberg model
\be\label{eq:HXXZ}
H_{\rm XXZ}=\frac{J}{4}\sum_\ell \sigma_\ell^x\sigma_{\ell+1}^x+\sigma_\ell^y\sigma_{\ell+1}^y+\Delta\sigma_\ell^z\sigma_{\ell+1}^z\, .
\ee 
This is the paradigm of (interacting) spin chains exactly solvable by algebraic Bethe ansatz~\cite{Korepinbook}. It is critical for $-1\leq \Delta\leq 1$ and the ground state is antiferromagnetic ($J>0$) for $\Delta>1$.

Recently, the time evolution under the Hamiltonian of the XXZ model has been a subject of intensive research~\cite{FE_13b,FCEC-13,Pozsgay:13a,bpgda-10, la-13}.
Here we show that at large $\Delta$ a time window is opened in which local degrees of freedom pre-relax to stationary values different from the  values (expected) in the generalized Gibbs ensemble.

The limit of large $\Delta$ is generally called `Ising limit', although at the large energy scales characteristic of global quenches the model is still interacting. This is a significant complication, which  will be circumvent for a particular class of initial states.

Let us first consider the time evolution of a generic state. The time evolution operator can be written as follows
\be
e^{-i H_{\rm XXZ } t}=e^{-i  \frac{\Delta J t}{4}\sum_\ell \sigma_\ell^z\sigma_{\ell+1}^z}\mathrm T\exp\Bigl(-i\int_0^t\mathrm d\tau  V_{\rm XXZ}(\tau)\Bigr)\, ,
\ee
where
\be
 V_{\rm XXZ}(\tau)= e^{i  \frac{\Delta J \tau}{4}\sum_\ell \sigma_\ell^z\sigma_{\ell+1}^z}H(0,0)e^{-i  \frac{\Delta J \tau}{4}\sum_\ell \sigma_\ell^z\sigma_{\ell+1}^z}
\ee
and we used the notations of \eref{eq:HXY} for the Hamiltonian of the XX model. Since the unitary transformation is very simple, $ V_{\rm XXZ}$ can be written explicitly:
\be\label{eq:Efft}
 V_{\rm XXZ}(\tau)= G+e^{i\Delta J \tau} F+e^{-i\Delta J \tau} F^\dag\, ,
\ee
where
\begin{eqnarray}
\fl \quad F=\frac{J}{16}\sum_{\ell}(\sigma_\ell^x+i\sigma_{\ell-1}^z\sigma_\ell^y)(\sigma_{\ell+1}^x+i\sigma_{\ell+1}^y\sigma_{\ell+2}^z)+(\sigma_\ell^y+i\sigma_{\ell-1}^z\sigma_\ell^x)(\sigma_{\ell+1}^y+i\sigma_{\ell+1}^x\sigma_{\ell+2}^z)\, ,\nn
\fl \quad G=\frac{J}{4}\sum_\ell\frac{1+\sigma_{\ell-1}^z\sigma_{\ell+2}^z}{2}(\sigma_\ell^x\sigma_{\ell+1}^x+\sigma_{\ell}^y\sigma_{\ell+1}^y)\, .
\end{eqnarray}
Notice that $G$ commutes with $\sum_\ell\sigma_\ell^z\sigma_{\ell+1}^z$ and is in fact the term that appears multiplied by $\Delta^2$ in the third conservation law of the XXZ model (where the Hamiltonian is the first) \cite{higherCL}.   
We single out the stationary term $G$
 \begin{eqnarray}
\qquad \mathrm T\exp\Bigl(-i\int_0^t\mathrm d\tau  V_{\rm XXZ}(\tau)\Bigr)=e^{- i G t} U_1(t)\nn
U_1(t)=\mathrm T\exp\Bigl(-i\int_0^t\mathrm d \tau e^{i\Delta\tau} e^{i G \tau}F e^{-i G \tau}+e^{-i\Delta\tau} e^{i G \tau} F^\dag e^{-i G \tau}\Bigr)\, .
 \end{eqnarray}
At fixed time, $ U_1(t)$ can be formally expanded in powers of $1/\Delta$ as follows
\be\label{eq:U1t}
U_1(t)=\exp\Bigl(-i \frac{H_1(t, e^{i\Delta t})}{\Delta}-i\frac{H_2(t,e^{i\Delta t})}{\Delta^2}-\cdots\Bigr)\, ,
\ee
where the Hermitian operators $H_j(x, y)$ do not depend explicitly on $\Delta$\footnote{This could be proved by induction, integrating by parts (choosing $e^{i n \Delta \tau}$ as the function to be integrated) a generic term of the series expansion of $U_1(t)$.}. Therefore, at fixed time and in the limit of large $\Delta$, at the leading order of perturbation theory, $U_1(t)$ can be replaced by the identity.
This is equivalent to approximate $H_{\rm XXZ}$ with the effective Hamiltonian
\be\label{eq:Heff0}
H_{\rm eff}=\frac{J}{4}\sum_\ell \frac{1+\sigma_{\ell-1}^z\sigma_{\ell+2}^z}{2}(\sigma_\ell^x\sigma_{\ell+1}^x+\sigma_{\ell}^y\sigma_{\ell+1}^y)+\Delta \sigma_\ell^z\sigma_{\ell+1}^z\, .
\ee
This is an \emph{interacting} translation invariant operator with much more symmetries than \eref{eq:HXXZ}; in particular:
\begin{enumerate}
\item there are local conservation laws that break one-site shift invariance, \emph{e.g.}
\be
\sum_\ell \sigma_{2\ell-1}^z\sigma_{2\ell}^z\qquad {\rm and}\qquad \sum_\ell \sigma_{2\ell}^z\sigma_{2\ell+1}^z\, ;
\ee 
\item \label{p:2}there are two invariant subspaces specified by the projectors
\be
{\rm P}_{\pm}=\prod_\ell\frac{1\pm \sigma^z_{2\ell-1}\sigma^z_{2\ell}}{2}\, .
\ee
\end{enumerate}

From the first property we realize that if the initial state breaks one-site shift invariance, translation invariance will not be restored at large times. However, this is only a property of the effective Hamiltonian; at late times correlation functions are expected to become translation invariant because of the contributions that we are neglecting.
 
Before investigating the consequences of the second property we prove it. Let $\ket{\varphi_\pm }$ a state that belongs to one of the two subspaces, namely $\sigma_{2\ell-1}^z\sigma_{2\ell}^z\ket{\varphi_\pm}=\pm \ket{\varphi_\pm}$ for any $\ell$ (the sign is independent of $\ell$). We have
\be\label{eq:sub}
\sigma_{2\ell-1}^z\sigma_{2\ell}^z H_{\rm eff}\ket{\varphi_\pm}=[\sigma_{2\ell-1}^z\sigma_{2\ell}^z, H_{\rm eff}]\ket{\varphi_\pm}\pm  H_{\rm eff}\ket{\varphi_\pm}\, .
\ee
The commutator is readily calculated
\be
i [\sigma_{2\ell-1}^z\sigma_{2\ell}^z, H_{\rm eff}]=D_{2\ell}(\sigma_{2\ell+2}^z+\sigma_{2\ell-1}^z)- D_{2\ell-2}(\sigma_{2\ell}^z+\sigma_{2\ell-3}^z)
\ee
where $D_\ell=\sigma_{\ell}^x\sigma_{\ell+1}^y-\sigma_{\ell}^y\sigma_{\ell+1}^x$. Since, for any $j$,
\begin{eqnarray}
(\sigma_{2j}^z+\sigma_{2j-3}^z)\ket{\varphi_{\pm}}=\pm(\sigma_{2j-1}^z+\sigma_{2j-2}^z)\ket{\varphi_{\pm}}\nn
D_{j}(\sigma_{j}^z+\sigma_{j+1}^z)=0\, ,
\end{eqnarray}
we obtain
\be
i [\sigma_{2\ell-1}^z\sigma_{2\ell}^z, H_{\rm eff}]\ket{\varphi_{\pm}}=0\, ,
\ee
that is to say, $H_{\rm eff}$ preserves the subspaces (\emph{cf.} \eref{eq:sub}), proving property~\eref{p:2}.

We now consider the time evolution of states of type $\ket{\varphi_\pm}$. 
Since the effective Hamiltonian acts as a block diagonal operator on the two subspaces, we can add to $H_{\rm eff}$ any Hermitian operator that is in the kernel of $P_{\pm}$ without affecting the time evolution.

The term of $H_{\rm eff}$~\eref{eq:Heff0} that is not multiplied by $\Delta$ is in the kernel of ${\rm P_+}$, therefore the evolution of states of type $\ket{\varphi_+}$ is simply generated by
\be
H^{(+)}_{\rm eff}=\frac{J\Delta}{4}\sum_\ell (\sigma_{2\ell}^z\sigma_{2\ell+1}^z+1)\, ,
\ee
which does not allow any form of nontrivial relaxation.

On the other hand, for states of type $\ket{\varphi_-}$ it is convenient to add the operator
\be
\frac{J}{4}\sum_\ell\frac{1-\sigma_{\ell-1}^z\sigma_{\ell-2}^z}{2}(\sigma_{\ell}^x\sigma_{\ell+1}^x-\sigma_{\ell}^y\sigma_{\ell+1}^y)\, ,
\ee
which is instead in the kernel of ${\rm P}_-$. This can be easily verified using the identity $\sigma_{\ell}^y\sigma_{\ell+1}^y=-\sigma_{\ell}^x\sigma_{\ell+1}^x\sigma_{\ell}^z\sigma_{\ell+1}^z$ and then applying the $\sigma^z$ matrices to $\ket{\varphi_-}$. 

We finally end up with the effective Hamiltonian
\be
H^{(-)}_{\rm eff}=\frac{J}{4}\sum_\ell \sigma_{\ell}^x\sigma_{\ell+1}^x+\sigma_{\ell-1}^z\sigma_{\ell}^y\sigma_{\ell+1}^y\sigma_{\ell+2}^z+\Delta \sigma_{\ell}^z\sigma_{\ell+1}^z\, .
\ee
The advantage of working with the latter Hamiltonian is evident: $H^{(-)}_{\rm eff}$ is a noninteracting operator in the Jordan-Wigner fermions with quantization axis along $y$ or, equivalently, in the Majorana fermions~\eref{eq:JW} up to a rotation about $x$
\begin{eqnarray}\label{eq:effXXZ}
H^{(-)}_{\rm eff}=e^{i \frac{\pi}{2} S^x }\overline H_{\rm eff}e^{-i \frac{\pi}{2} S^x }\nn
\overline H_{\rm eff}=\frac{J}{4}\sum_\ell \sigma_{\ell}^x\sigma_{\ell+1}^x+\sigma_{\ell-1}^y\sigma_{\ell}^z\sigma_{\ell+1}^z\sigma_{\ell+2}^y+\Delta \sigma_{\ell}^y\sigma_{\ell+1}^y\, ,
\end{eqnarray}
where $S^x=\frac{1}{2}\sum_\ell\sigma_\ell^x$. 

We note that the operator $\overline H_{\rm eff}$ can be written in terms of the local charges of the XY Hamiltonian \eref{eq:HXY0} in the (Ising) limit $\gamma\rightarrow-1$ (\emph{cf}. \eref{eq:I}):
\be
\overline{H}_{\rm eff}=\frac{\Delta}{2}I_1^{+(e)}+I_2^{+(e)}\, .
\ee
As a consequence, the effective Hamiltonian has the same `oversized' set of local conservation laws of the XY model \eref{eq:Q}\eref{eq:I}, signaling that for large $\Delta$ the time evolution can experience pre-relaxation.

\subsection{Quench from the ground state of the Majumdar-Ghosh model} 
We consider again a quench from the Majumdar-Ghosh dimer product state~\eref{eq:MGs}, which is of type $\ket{\varphi_-}$. The state is invariant under rotations $e^{-i\frac{\pi}{2} S^x}\ket{\Psi_0}=\ket{\Psi_0}$ so, within our approximation, the time evolution of the expectation value of a generic operator $\mathcal O$ reads as
\be
\fl\qquad\qquad\braket{\Psi_0|e^{i H_{\rm XXZ} t}\mathcal Oe^{-i H_{\rm XXZ} t}|\Psi_0}\approx \braket{\Psi_0|e^{i \overline H_{\rm eff} t}(e^{-i \frac{\pi}{2} S^x }\mathcal O e^{i \frac{\pi}{2} S^x } )e^{-i \overline H_{\rm eff} t}|\Psi_0}\, .
\ee
Expectation values are therefore completely determined by the correlation matrix $\Gamma(t)$ of $e^{-i \overline H_{\rm eff} t}\ket{\Psi_0}\bra{\Psi_0}e^{i \overline H_{\rm eff} t}$.

The symbol of the two-site representation of $\overline H_{\rm eff}$~\eref{eq:effXXZ} is readily obtained 
\be\label{eq:symbolXXZ}
\overline{\mathcal H}_{\rm eff}(k)=-\overline{\varepsilon}(k)\tau_1^x e^{i\frac{k}{2}\tau_1^z}\tau_2^y e^{-i\frac{k}{2}\tau_2^z}\, ,
\ee
where $\overline{\varepsilon}(k)=J (\cos k+\frac{\Delta}{2})$.
The corresponding time evolution matrix is given by
\be
e^{-i\overline{\mathcal H}_{\rm eff}(k) t}=\cos(\overline\varepsilon(k)t)+i \sin(\overline\varepsilon(k)t)\tau_1^x e^{i\frac{k}{2}\tau_1^z}\tau_2^y e^{-i\frac{k}{2}\tau_2^z}
\ee
We finally find (the correlation matrix in the initial state was reported in \eref{eq:MGcm})
\begin{eqnarray}\label{eq:predXXZ}
\fl\qquad\Gamma(t,k)=&e^{-i\overline{\mathcal H}_{\rm eff}(k) t}\tau_1^x \tau_2^ye^{i\overline{\mathcal H}_{\rm eff}(k) t}=\tau_1^x\tau_2^y+2\sin k\sin(\overline\varepsilon(k)t)\Bigl\{\cos(\overline\varepsilon(k)t)\frac{\tau_1^z-\tau_2^z}{2}\nn
\fl\qquad & +\cos k\sin(\overline\varepsilon(k)t)\frac{\tau_1^x\tau_2^x+\tau_1^y\tau_2^y}{2}+\sin k\sin(\overline\varepsilon(k)t)\frac{\tau_1^y\tau_2^x-\tau_1^x\tau_2^y}{2}\Bigr\}\, .
\end{eqnarray}
From this expression we can immediately extract the symbol of the correlation matrix of the generalized Gibbs ensemble that describes pre-relaxation (again, removing the rapidly oscillating terms):
\be\label{eq:preXXZ}
\Gamma_{\rm pGGE}(k)=\tau_1^x\tau_2^y+\sin(2k)\frac{\tau_1^x\tau_2^x+\tau_1^y\tau_2^y}{4}+\sin^2 k\frac{\tau_1^y\tau_2^x-\tau_1^x\tau_2^y}{2}\, .
\ee
At this order of approximation the correlation matrix at large times is independent of $\Delta$, exactly as in the quantum XY model \eref{eq:corr0} it was independent of the magnetic field. 

\paragraph{On a noninteracting description after pre-relaxation. }
It is important to note that we have been able to reduce the problem to a noninteracting one only because the effective Hamiltonian $H_{\rm eff}$ \eref{eq:Heff0} preserves the subspace specified by ${\rm P}_-$. 
Using the noninteracting effective Hamiltonian $H^{(-)}_{\rm eff}$ as a starting point for describing the time evolution following pre-relaxation is therefore not safe.

In the following we provide a simple argument against the possibility to formulate a noninteracting description of the subsequent dynamics based on $H^{(-)}_{\rm eff}$.

Following the procedure described in \Sref{s:time}, because translation invariance is supposed to be eventually restored, at late times the perturbation to the superintegrable model should select the maximal set of translation invariant local conservation laws in involution. This observation allows us to construct the generalized Gibbs ensemble without any knowledge of the time-dependent GGE. 
As a matter of fact, we can use the result already obtained for the XY model (the first equation of \eref{eq:GammaGGE}), obtaining a gaussian prediction for the GGE. Substituting $\Gamma_{\rm pGGE}(k)$ \eref{eq:preXXZ} into \eref{eq:GammaGGE} gives the symbol of the correlation matrix
\be\label{eq:XXZGGEf}
\Gamma^{free}_{\rm GGE}(k)=\frac{1+\cos k}{2}\tau_1^x e^{i\frac{k}{2}\tau_1^z}\tau_2^y e^{-i\frac{k}{2}\tau_2^z}\, .
\ee 
By construction, translation invariance is restored, however the correlation matrix with symbol~\eref{eq:XXZGGEf} has a considerable problem: the $U(1)$ symmetry of rotations about $z$ has been lost. This can be realized by computing the nearest neighbor correlators:
\bea
\braket{\sigma_1^z\sigma_2^z}^{free}_{\rm GGE}=i\braket{a_1^x a_2^y}=-\frac{1}{2}\nn
\braket{\sigma_1^x\sigma_2^x}^{free}_{\rm GGE}=-i\braket{a_1^y a_2^x}=-\frac{1}{4}\nn
\braket{\sigma_1^y\sigma_2^y}^{free}_{\rm GGE}=-\braket{a_1^y a_1^x a_2^y a_2^x}=-\frac{1}{8}\, .
\eea
Since $\braket{\sigma_1^x\sigma_2^x}^{free}_{\rm GGE}\neq \braket{\sigma_1^y\sigma_2^y}^{free}_{\rm GGE}$, the GGE and, in turn, the time-dependent GGE can not be gaussian in the (noninteracting) fermions that diagonalize $H^{(-)}_{\rm eff}$!

\subsection{Numerical results}
Using Wick theorem and \eref{eq:predXXZ} we can compute the correlation functions of spin operators.
In order to compare the approximate results with the numerical data obtained in \cite{FCEC-13}, we focus on  short-range correlators. 
Within our approximation, many correlators are constant because of the symmetries of the correlation matrix~\eref{eq:predXXZ}; $\braket{\sigma_\ell^x\sigma_{\ell+1}^x}$ is one of the correlators with  a nontrivial time evolution:
\be\label{eq:sxsx}
\braket{\sigma_\ell^x\sigma_{\ell+1}^x}\sim\left\{\begin{array}{cl}
0&\ell \ {\rm even}\\
-\frac{1}{2}-\int_{-\pi}^\pi\frac{\mathrm d k}{2 \pi} \sin^2 k \cos[(\Delta+2\cos k)t]&\ell \ {\rm odd}\, .
\end{array}\right.
\ee
Since we neglected $O(1/\Delta)$ contributions (\emph{cf}. \eref{eq:U1t}), we expect $O(1/\Delta)$ corrections to the expectation values (at fixed time). 

\begin{figure}[htbp]
\begin{center}
\includegraphics[width=0.49\textwidth]{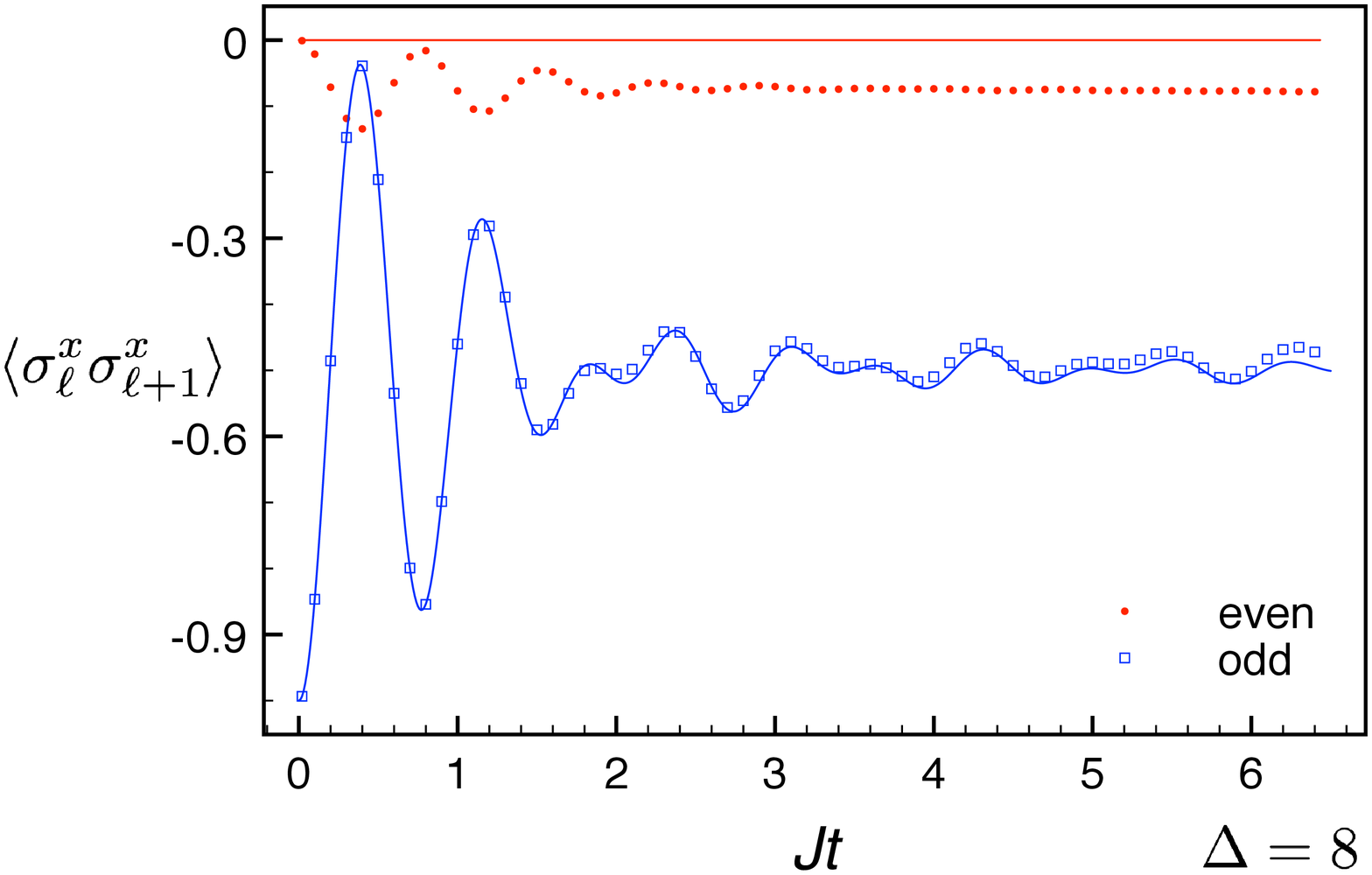}
\includegraphics[width=0.49\textwidth]{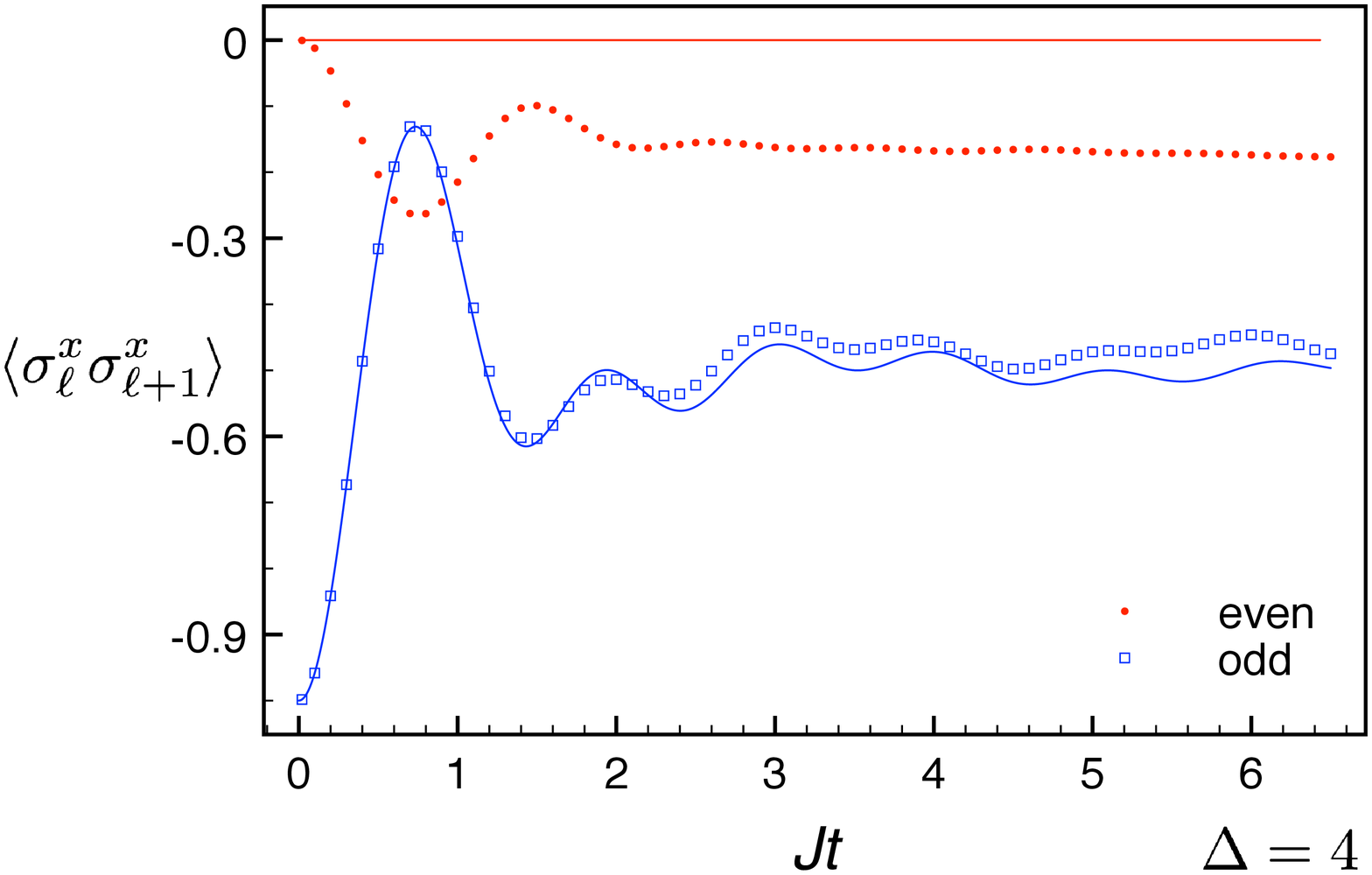}
\caption{Nearest neighbor correlators $\braket{\sigma_\ell^x \sigma_{\ell+1}^x}$ after a quench from the Majumdar Ghosh ground state in the XXZ model with $\Delta=8$ (left) and $\Delta=4$ (right). 
The points are tDMRG data for an (open) chain of 32 sites~\cite{FCEC-13}. 
The solid lines represent the time evolution \eref{eq:sxsx} under the (noninteracting) effective Hamiltonian $H_{\rm eff}^{(-)}$ \eref{eq:effXXZ}. In the time window considered there are corrections $O(1/\Delta)$.
}
\label{fig:XXZ}
\end{center}
\end{figure}

\Fref{fig:XXZ} shows the results for two quenches with rather large $\Delta$.
In the time window considered,  $\braket{\sigma_1^x\sigma_2^x}$ is in very good agreement with \eref{eq:sxsx}, instead the $O(1/\Delta)$ corrections to $\braket{\sigma_2^x\sigma_3^x}$ are clearly visible. 

\section{Conclusions}\label{s:conc}
We have shown that some extensively studied models have extra families of local charges, in addition to the translation invariant ones that are usually taken into account. We presented a systematic method to study the $n$-site shift invariant conservation laws of noninteracting models and  constructed the two-site shift invariant charges of the quantum XY model without magnetic field.

We investigated the quench dynamics in the presence of a small perturbation that breaks the hidden symmetries underlying the additional conservation laws. We found that sufficiently small subsystems (much smaller than the characteristic length introduced by the perturbation) experience pre-relaxation. The almost stationary properties can then be described in terms of a generalized Gibbs ensemble constructed with the local conservation laws of the  unperturbed model.  

A quite remarkable result is that even the relaxation process following pre-relaxation can be described by a generalized Gibbs ensemble, which is however time dependent. 
We checked our analytic results against numerics for the non-equilibrium evolution under the Hamiltonian of the quantum XY model with a small magnetic field. 

We finally shown that this type of pre-relaxation is not peculiar of noninteracting models. We indeed established that the slow restoration of translation invariance observed in \cite{FCEC-13} after quenches in the XXZ spin-1/2 chain for large anisotropy $\Delta$ finds a natural interpretation in our construction. In particular, we solved the dynamics at the leading order in $1/\Delta$ for a quench from the Majumdar-Ghosh dimer product state, showing that translation invariance is not restored at the large times at which the most local observables pre-relax.

Our analysis raises many interesting questions:
\begin{itemize}
\item[-] We have not investigated whether particular symmetries of the one-particle spectrum in noninteracting models could result in interacting local conservation laws.
Because of the relation between local charges and generalized Gibbs ensemble, the analysis of the time evolution from initial states that are not Slater determinants could be useful to address this issue.
\item[-] The pre-relaxation behavior after quenches close to superintegrable points is presumably captured by the time evolution of the entanglement entropy of subsystems (of intermediate length). 
It could be worth studying whether in the scaling limit in which the time-dependent GGE is defined the entanglement entropy displays some `universal' behavior (in the sense of \cite{cc-05,FC:2008}). 
\item[-] The type of pre-relaxation discussed in this paper is strongly dependent on the ratio between the typical length of the observable under investigation and the typical length introduced by the perturbation. A scaling analysis of correlation functions is therefore the next step towards the characterization of quench dynamics in these models.
\item[-] Concerning quantum quenches in the XXZ model, there are many open questions. One of the most relevant is how the quasi-local conservation laws that have been recently constructed~\cite{prosen} enter into the definition of the generalized Gibbs ensemble. In the light of our results, we wonder whether the set of independent (quasi-)local charges of the XXZ model is larger than a maximal set of local conservation laws in involution, as in the quantum XY model (with zero magnetic field). 
\item[-] The effective noninteracting Hamiltonian that describes the time evolution of the Majumdar-Ghosh dimer product state in the XXZ model is equivalent to the Ising limit of the XXZ Hamiltonian only in a tiny subspace of the Hilbert space (to which the initial state belongs). We constructed a generalized Gibbs ensemble in terms of the local conservation laws of the noninteracting model, but we have not established the relation with the (quasi-)local charges of the original model. This is however a fundamental step to demonstrate that the stationary state is fully characterized by the local conservation laws. 
\end{itemize}
A final remark. A time-dependent GGE can be also used to describe the time evolution of local observables under more complicated protocols in which the Hamiltonian of the superintegrable model is perturbed by some charges (not commuting with one another) with time dependent coupling constants. 

\ack
This work was supported by the EPSRC under grants EP/I032487/1 and EP/J014885/1.
The tDMRG data of \Fref{fig:XXZ} were obtained by Mario Collura during the preparation of \cite{FCEC-13}.  
I thank Claudio Bonati, Pasquale Calabrese, Mario Collura, and Fabian Essler for stimulating discussions.

\end{document}